# Structure without Law: Structural Nonrealism and Quantum Information

*Arkady Plotnitsky\**

**Abstract.** The article introduces a new concept of structure, defined, echoing J. A. Wheeler's concept of "law without law," as a "structure without law," and a new philosophical viewpoint, that of structural nonrealism, and considers how this concept and this viewpoint work in quantum theory in general and quantum information theory in particular. It takes as its historical point of departure W. Heisenberg's discovery of quantum mechanics, which, the article argues, could, in retrospect, be considered in quantum-informational terms, while, conversely, quantum information theory could be seen in Heisenbergian terms. The article takes advantage of the circumstance that any instance of quantum information is a "structure"—an organization of elements, ultimately bits, of classical information, manifested in measuring instruments. While, however, this organization can, along with the observed behavior of measuring instruments, be described by means of classical physics, it cannot be predicted by means of classical physics, but only, probabilistically or statistically, by means of quantum mechanics, or in high-energy physics, by means of quantum field theory (or possibly some alternative theories within each scope). By contrast, the emergence of this information and of this structure cannot, in the present view, be described by either classical or quantum theory, or possibly by any other means, which leads to the concept of "structure without law" and the viewpoint of structural nonrealism. The article also considers, from this perspective, some recent work in quantum information theory.

**Key words**: causality, nonrealism, quantum information, quantum theory, reality, realism, structure

## 1. Introduction

The article introduces a new concept of structure, defined, echoing J. A. Wheeler's concept of "law without law," as a "structure without law," and a new philosophical viewpoint, that of structural nonrealism, and considers how this concept and this viewpoint work in quantum theory in general and quantum information theory in particular. It takes as its historical point of departure W. Heisenberg's thinking leading him to his discovery of quantum mechanics, as matrix mechanics. I shall argue that, while this thinking could not be rigorously claimed to be quantum-informational, it could be viewed as quantum-informational *in spirit*, and conversely, quantum information theory could be viewed as Heisenbergian in spirit.[1] N. Bohr, in his initial assessment of Heisenberg's discovery in 1925 (before E. Schrödinger's introduction of his wave version), described Heisenberg's thinking and his "new quantum mechanics" as follows:

---

[1] This view was initially proposed by the present author in (Plotnitsky 2002) and developed in (Plotnitsky 2016), without, however, casting it in terms of structural nonrealism.

\*Theory and Cultural Studies Program, Purdue University, W. Lafayette, IN, 47907, USA

email: plotnits@purdue.edu

> In contrast to ordinary mechanics, *the new quantum mechanics does not deal with a space–time description of the motion of atomic particles*. It operates with manifolds of quantities [matrices] which replace the harmonic oscillating components of the motion and symbolize the possibilities of transitions between stationary states in conformity with the correspondence principle [which requires that quantum and classical predictions coincide in the classical limit]. These quantities satisfy certain relations which take the place of the mechanical equations of motion and the quantization rules [of the old quantum theory]. (Bohr 1987, v. 1, p. 48; emphasis added)

This was a radical departure from the preceding history of modern physics, from Galileo's mechanics to A. Einstein's relativity and even to the previous ("old") quantum theory. All these theories were based on such descriptions or representations, either phenomenally visualizable, at least in dealing with elemental individual processes, as, for example, in classical mechanics, or beyond visualization but given a conceptual or mathematical representation, as, for example, in relativity when dealing with photons and velocities close to $c$. While we can give the relativistic behavior of photons a concept expressible in language and represent it mathematically, we have no phenomenal means of visualizing this behavior, or the behavior represented by Einstein's velocity-addition formula. This was the first physical theory that defeated our ability to form a phenomenal conception of an elemental physical process, and, as such, it was in turn a radical change in the history of physics. The reason that it worried Einstein less than the situation that emerged as regards the behavior of the elemental constituents of nature in quantum physics was that relativity still offered a conceptual-mathematical representation (causal in character) of the relativistic behavior in question, an aspect of Einstein's view that I shall consider below. Ultimately, photons are quantum objects and are treated by quantum electrodynamics, which, in the view adopted here, no longer represents the behavior of quantum objects in the corresponding (high-energy) regimes even mathematically, any more than quantum mechanics does. Classical statistical physics still relied and indeed depended on the representational treatment of the behavior of the elemental constituents of the multiplicities considered, such as the molecules of a gas, constituents that were viewed as behaving in accordance with the laws of classical mechanics. The description of individual quantum objects and the corresponding mathematical representation became partial in the so-called old quantum theory, such as Bohr's atomic theory, introduced in 1913 and developed by him and others over the following decade. The old quantum theory only provided such a description and mathematical representation, in terms of orbits, for stationary states of electrons in atoms, but not for the transitions, "quantum jumps," between stationary states. For that reason, the old quantum theory was called "semi-classical." Heisenberg, as Bohr's statement just cited explained, abandoned a geometrical representation of stationary states as well and, thus, any description or representation, even a mathematical one, of *the behavior of quantum objects*. In his scheme, stationary states were only represented by energy-level values, considered apart from representing the behavior of electrons themselves to which these energy levels were assigned. The theory thus only was only concerned with transition probabilities between physical states, indeed between events observed in measuring instruments impacted by quantum objects, and not with the physical behavior of quantum objects, such as electrons, in space and time, represented along geometrical lines, as in classical physics or relativity. As such, Heisenberg's theory may also be seen in terms of transitions from geometry to algebra in fundamental physics, which compelled Einstein to speak of Heisenberg's "purely algebraic method." As he said: "perhaps the success of the Heisenberg method points to a purely algebraic method of description of nature, that is, to the elimination of continuous functions from physics [as the way of representing physical reality]. Then, however, we must give up, in principle, the space–time continuum" (Einstein 1936, p. 378). (Einstein 1936, p. 378). I shall return to this aspect of Heisenberg's discovery in Section 4.

For the moment, Heisenberg's approach may be thought of in quantum-informational terms because the quantum-mechanical situation, as he conceived of it (initially dealing with hydrogen spectra), was in effect defined by:

(a) certain *already obtained* information, concerning the energy of an electron, derived from spectral lines (due to the emission of radiation by the electron), *observed* in measuring instruments; and

(b) certain possible future information, concerning the energy of this electron, *to be obtainable* from spectral lines *to be observed* in measuring instruments and predictable, unavoidably (on experimental grounds) in probabilistic or statistical terms, by means of the mathematical formalism of one or another quantum theory.

Heisenberg's strategy was to develop a mathematical formalism that would connect these two sets of data, manifested in measuring instruments, only in predictive terms, moreover (in accord with what is actually observed in quantum experiments), in strictly probabilistically or statistically predictive terms, without assuming that this formalism needed to represent how these two sets of data or information are connected by a spatiotemporal process or how each set comes about, in the first place. This type of representation or the physical conception of the processes that would be thus represented mathematically (again, still possible in Bohr's atomic theory in the case of stationary states) did not appear possible at the time and has not been possible since, at least not in a way generally agreed upon. Heisenberg's mathematical scheme did not represent anything at the time of measurement either: it only predicted transition probabilities between situations defined by measurements, those already performed, which provide the numerical data that serve as the experimental basis for these predictions, and possible future ones.

Part of the mathematical structure of Heisenberg's scheme was provided by the equations of classical mechanics, directly borrowed by Heisenberg in accordance with the correspondence principle, suitably modified by Heisenberg, as explained in Section 4. Classical variables, however, would only give correct predictions in the classical limit, for example, in the case of large quantum numbers for electrons, where their behavior could be treated classically, even though the corresponding processes would still need to be assumed to be quantum. To remedy this failure, Heisenberg replaced classical variables with new variables, of a type never used in physics previously. They were complex-valued (infinite-dimensional) matrix variables, essentially Hilbert-space operators, in J. von Neumann's more rigorous version of the formalism, introduced shortly thereafter, and commonly used ever since.

Heisenberg's approach was quickly extended to other quantum predictions, all in general probabilistic or statistical in nature. Any such prediction was, again, concerned with a future measurement and the information it would provide, the information manifested only as an effect of the interaction between the quantum object considered and a suitable measuring apparatus, on the basis of a previously performed measurement and the information obtained from it, manifested as the same type of effect.

One was, thus, no longer concerned, as in classical physics or relativity, with predicting, even in the case of elemental individual processes, the space-time behavior of the objects considered due to (continuous) changes in their states, assumed to be definable independently of the interactions between objects and measuring instruments, and representable by the corresponding mathematical formalism. Instead, following Bohr's thinking in his 1913 atomic theory and related approaches found in the old quantum theory, one was thinking in terms of discontinuous transitions between physical states of quantum objects and in terms of predicting the probabilities or statistics of such transitions, as only probabilistic predictions were possible experimentally.[2] These states are, moreover, only manifested as effects of the interactions between quantum objects and measuring instruments. While not part of Bohr's thinking or at least not part of his argumentation in his 1913 theory, this understanding became central to his thinking following Heisenberg's discovery of quantum mechanics and came to define Bohr's interpretation of quantum mechanics and quantum phenomena themselves.[3]

---

[2] I am indebted to L. Freidel for this observation, discussed in (Freidel 2016).

[3] Although quantum phenomena and quantum mechanics are commonly interpreted jointly within the same framework, as they are in Bohr or in this article, these are separate entities, each of which could be given its own interpretation. Besides, quantum phenomena in the same interpretation, could be handled by different theories.

This article will explore the implications of quantum-theoretical thinking, from Heisenberg's quantum mechanics to quantum information theory, for the idea of structure in physics and beyond, most especially, as announced from the outset, the concept of structure without law and the viewpoint of structural nonrealism, which are this article's main contributions. A qualification is in order before I proceed.

The argument of this article belongs to the philosophy of physics. It represents, however, a different form of philosophy of physics, vis-à-vis most other forms of the institutional philosophy of physics, specifically the philosophy of quantum theory, apart from some more historically oriented studies, where some aspects of the present approach could occasionally be found. This difference is especially reflected in my emphasis on *thinking* concerning physics, which emphasis defines the philosophy of physics in question. It is the philosophy of thinking in fundamental physics, by which I mean those areas of experimental and theoretical physics that are concerned with the ultimate constitution of nature, as we, as human beings, understand this constitution. I qualify because, in the view adopted by this article, this constitution could only be assumed as something conceived by the human mind or something assumed to be beyond human conception. The thinking in questions refers to both thinking by the key figures considered here and our own thinking, that of this article's readers included. It may, it is true, happen that the same theory results from a different way of thinking, as was in fact the case in Heisenberg's nonrealist thinking and Schrödinger's realist thinking that, nevertheless, led each of them to quantum mechanics. This difference is diminished by the fact that Schrödinger could not entirely avoid some of the quantum principles used by Heisenberg, principles that resisted realism and, as a result, created tensions within Schrödinger's scheme, which he was ultimately unable to resolve. In any event, if anything, this potential for different ways of thinking to arrive at the same theory makes this approach to the philosophy of physics all the more relevant.

It is important, however, that this philosophical argument does involve physics. Indeed, it is as much about physics as about the philosophy of physics, by definition, because this argument concerns *thinking about physics*. This fact is often missed in dealing with certain philosophically oriented arguments, as it has often been in response to Bohr's work, sometimes alleged not to have involved any new physics. Nothing could be further from the truth. It is true that Bohr's work did not offer any new mathematics of quantum theory or new experimental findings, and the lack of both, especially the former, has sometimes served to claim that there was no new physics either. But that does not mean that there was no new physics in his argumentation. Bohr's interpretation of quantum phenomena and quantum mechanics was also new physics.

I shall proceed as follows. The next section gives a general introduction to structural nonrealism. Section 3, first, outlines the fundamental concepts used by it, such as "reality," "theory," "model," "structure," "causality," "complementarity," and the concept of concept itself, and then addresses some of the key physical and philosophical implications of structural nonrealism. Section 4 revisits Heisenberg's discovery of quantum mechanics as a program of advancing from fundamental physical principles to mathematical structures, in the absence of the space-time description of the behavior of quantum objects, an approach that anticipated, correlatively, both structural nonrealism and quantum informational thinking. As such, it also prepared Bohr's interpretation of quantum phenomena and quantum mechanics, considered in this section as well. Section 5 offers a discussion of certain parallel ways of thinking in some recent approaches to quantum information theory.

## 2. From Quantum Information to Structural Nonrealism

This article aims to give a conceptual grounding to, and to explore the implications of, the fact that *quantum information* (a concept that shall be applied here to essentially all data in question in quantum physics) is defined by a special set of *structures of physically classical information*, gathered from the data found in measuring instruments impacted by quantum objects, the existence of which is in fact inferred from this structure. By "structure" I refer to the *character*, sometimes organized and sometimes indeterminate or random (I shall explain these terms and differences between them below, treating them as interchangeable for now), of units (bits) of information, which are effects of the interactions between quantum objects and measuring instruments, effects manifested, as the data associated with quantum events, in the observable parts of these instruments. Using the term "structure" in this sense may be unorthodox. The term more customarily refers to a *strictly organized* collection of elements and the rule-governed relationships among them, such as a group in mathematics, and it will be used in this sense here as well, in particular when referring to mathematical structures, such as those defining the formalisms of physical theories. When, however, it comes to quantum data or information, the structures in question are, first, only organized in the sense of being statistically correlated. Secondly, they are so correlated only in certain specified experimental circumstances and are indeterminate or random in certain other specified experimental circumstances, always mutually exclusive with those of the first type, while, importantly, always allowing us to choose either one corresponding experimental setup or the other. This mutual exclusivity is a manifestation of Bohr's concept of complementarity, explained below. The *structures* of quantum information thus combine indeterminacy or randomness and a statistically correlated organization of a particular type, not found in classical physics or relativity, and hereafter referred to as "quantum correlations." Finally, and arguably most characteristically and most crucially, even quantum correlations still involve indeterminacy or randomness, because these correlations only pertain to certain collectivities of events, registered in the corresponding experimental devices, while each individual event involved remains an independent event as concerns our predictions about it.[4] It is a matter of interpretation whether such an event could be assigned, say, on Bayesian lines, a probability of occurrence and thus is merely indeterminate, or cannot be assigned such a probability and is thus entirely random (in the present definition of "indeterminacy" and "randomness," discussed in more detail below), as opposed to the statistics of multiple repeated, identically prepared, experiments. Such statistics can always be assigned and are well confirmed experimentally. The interpretation adopted here assumes the (strict) randomness of individual quantum events, while, however, keeping in mind the Bayesian view as a possible alternative. Quantum correlations themselves are collective, statistical, and as such they would not depend on either interpretation of our predictions concerning the individual events involved. The term "indeterminacy" (of our predictions concerning individual quantum events) will be used here to apply to both types of events and, thus, to either interpretation. Either way, that, in certain circumstances, indeterminate or random individual events form statistically correlated and thus statistically ordered multiplicities is one of the greatest mysteries, if not the greatest mystery, of quantum physics, which makes it as much about order as about indeterminacy and randomness. The order in question is, again, a statistically correlated order, or structure, without assuming the existence of an ontologically underlying classical order, or structure, that merely cannot be accessed epistemologically.

Either in its organization or in its indeterminacy or randomness, and, in the first place, in its constitutive elements (each of which defines the quantum phenomenon associated with a quantum event), such a structure can, in each situation, be *phenomenally* represented by means of classical physics. Physically, quantum phenomena are defined by the fact that, in considering them, Planck's constant, $h$, must be taken into account, which still allows one to use classical physics in describing them, although, again, not in predicting them. Quantum phenomena could only be predicted, thus far probabilistically or statistically (it is a matter of debate whether this indeterminism precludes some underlying causality), by means of an alternative, quantum, theory, which would inevitably contain $h$. The currently standard version of quantum theory, the only one to be considered in this article, is comprised of three theories.[5] The first, discovered first as well, is quantum mechanics for continuous variables in infinite-dimensional Hilbert-spaces (QM), the second is quantum theory for discrete variables in finite-dimensional Hilbert space (QTFD), and the third is quantum field theory in Hilbert spaces that are tensor products of finite and infinite dimensional Hilbert space (QFT). Such tensor products are also used in QM when one needs to incorporate spin, which was first done by W. Pauli (Pauli 1927). QFT, which handles high-energy physics, is currently comprised of several theories, jointly constituting the so-called standard model of particle physics: quantum electrodynamics (QED), the quantum field theory of weak forces, and the quantum field theory of strong forces, known as quantum chromodynamics (QCD). While the first two are unified or (as some prefer to see it) "merged" within the electroweak theory, the unification or merger of all three, sometimes referred to as "grand unification," has not been accomplished. Several group-theoretical models have been proposed, but none appears to be sufficiently satisfactory to be commonly accepted. Besides, achieving this unification is not always seen, including by this author, as imperative, insofar as the set of theories involved predict all the available data in their respective domains; there is, thus far, no experimental evidence that a grand unified theory of one kind or another is necessary. The capacity of QFT to quantize gravity is a different matter, because QFT and general relativity, the currently standard theory of gravity, are inconsistent with each other. This inconsistency is arguably the greatest outstanding problem of fundamental physics, which motivated string and then M-brane theories, and several alternative proposals, including in quantum information theory (e.g., Hardy 2007).

---

[4] The situation becomes more nuanced in the case of the famous thought experiment proposed by A. Einstein, B. Podolsky, and N. Rosen, the EPR experiment (Einstein et al 1935), discussed below, because in this case we deal with statistical correlations pertaining to multiple paired events. These nuances, however, do not affect my point at the moment, as explained in detail in (Plotnitsky 2016, pp. 136-154).

[5] Alternative theories of quantum phenomena, such as Bohmian mechanics, will only be mentioned in passing here. For convenience, the term quantum theory, too, will refer here to the standard versions of the three theories just described, although, technically, by virtue of dealing with quantum phenomena, such alternative theories are quantum theories as well.

The interpretation of quantum phenomena and quantum theory adopted here, grounded in the viewpoint of "structural nonrealism," takes a nonrealist view of both, a view that follows Heisenberg and Bohr, and, with them, "the Copenhagen spirit of quantum theory" [*Kopenhagener Geist der Quantenheorie*], as Heisenberg called it (Heisenberg 1930, p. iv).[6] In most of this article, I will be concerned with QM, although I will give some attention to QTFD in the context of quantum information theory, which has, thus far, been primarily concerned with QTFD. QFT will only be mentioned in passing, although structural nonrealism fully applies in this case, and historically, QFT, beginning with QED, has been used to support the spirit of Copenhagen all along. I shall now outline the viewpoint of structural nonrealism and the corresponding interpretation of quantum phenomena and QM.

It is worth recalling, first, that classical physics, specifically classical mechanics and classical electrodynamics, or relativity predicts the phenomena it considers *because* it offers an idealized representation of these phenomena and processes connecting them by means of mathematical models. In classical physics, these processes are assumed to be

---

[6] This characterization, abbreviated here to "the spirit of Copenhagen," is preferable to that of the Copenhagen interpretation, because there is no single such interpretation, even in the case of Bohr, who changed his views a few times, sometimes significantly.

*causal*, and predictions concerning them are ideally exact, *deterministic*, in the case of individual classical objects or small classical systems, handled by classical mechanics or classical electrodynamics. (Such theories are sometimes referred to as value-definitive.) I distinguish "causality" and "determinism," referring by causality, ontologically, to the conception that the state of the system considered is determined at all future moments of time, once it is determined at a given moment of time, and by determinism, epistemologically, to the possibility of predicting the outcomes of such processes ideally exactly, although both concepts are connected and, at bottom, are co-defining in classical physics. This conception of causality, which defined physics since Galileo and Newton, or even Aristotle, and philosophy beginning at least with Plato, will be termed here "classical causality." As will be seen, causality may be defined differently, first, in a relativistic or local sense (although relativity itself is a classically causal and indeed deterministic theory) and, second, in a quantum-theoretical probabilistic or statistical sense. All determinism, as here defined, is classical. There are, however, alternative definitions of determinism as well, including in QM, in particular referring to Schrödinger's equation.

Probabilistic or statistical classical theories, such as classical statistical physics or chaos theory, do introduce certain complications into the physics governed by classical causality. These complications are, however, not essential because the behavior of these systems is assumed to be at bottom classically causal and governed by classical mechanics, which makes these limitations merely a practical matter due to the mechanical complexity of these systems. This assumption grounds classical statistical physics and is well justified within its scope. Indeed, one could experimentally isolate, say, a molecule of a gas and make deterministic predictions concerning its behavior. Quantum aspects of its behavior could be neglected within the scope of classical statistical physics without affecting such predictions.

The situation is fundamentally different in QM or QTFD and QFT, if one adopts a nonrealist, in-the-spirit-of-Copenhagen, view of these theories and of quantum phenomena themselves. While the *physical content* of all information obtainable from quantum phenomena can be physically represented by means of classical physics, the objects and processes, *quantum* in nature, responsible for the *emergence* of these phenomena and, hence, of the architecture of quantum information cannot be represented by means of QM, or conceivably by any other means, mathematical or not. This article takes an even stronger view, according to which, quantum objects and processes are not only beyond representation but are also, more radically, beyond conception: they are beyond the reach of thought altogether. Not all interpretations in the spirit of Copenhagen go that far. Thus, while there are indications in his writings that Bohr might have agreed with this view, he never expressly stated so. One could claim with more definitiveness that his argumentation implies that, even if such a conception is in principle possible, it could not be unambiguously used in in quantum theory (QM, QFT, QTFD). While this is a less stringent prohibition, this difference is only essential philosophically and does not affect the physics involved. This is because, even though such a conception is in principle *possible* in this less stringent view, no conception of the ultimate nature of quantum objects and processes is *available* in either view, as things stand now, a crucial qualification that applies to the present position as well and is assumed throughout this article. I shall explain why this qualification is applicable even if one assumes that no conception concerning quantum objects and processes is possible at all, now or ever, and comment on other nuances involved in this situation in more detail below, also as concerns the relationships between the physical and mathematical conceptions involved. One could assume, and many do, the possibility of a mathematical representation of quantum objects and processes in the absence of any physical conception of them. While Bohr's and the present interpretations of quantum phenomena and QM exclude this possibility as well, Heisenberg, appears to have been more open to it, in his later writings, along the lines of structural realism, a viewpoint discussed below.

The probabilistic or statistical nature of quantum predictions, in nonrealist interpretations, in contrast to classical mechanics or electrodynamics, even in the case of elemental individual quantum processes, is unavoidable, because of the absence of not only determinism, which is an experimentally established fact, but also classical causality. This absence is automatic under either assumption: that of the impossibility of a representation, conceptual or mathematical or that of the impossibility of a conception of quantum objects and behavior. If no representation, let alone conception, of quantum processes is available it would be difficult to think of them as causal, which would imply at least a partial representation and certainly a partial conception of them.

This does not preclude the possibility that the ultimate nature of the processes responsible for these phenomena (perhaps at levels beyond those currently covered by quantum theory in any of its forms, say, levels closer to Planck's scale than these theories are) is classically causal or could, in the first place, be represented by the corresponding theory. For one thing, there is no way to be certain that any (apparently) indeterminate or random event or sequence of events is in fact indeterminate or random, rather than causally connected, and there is no proof in the mathematical theories dealing with indeterminacy and randomness that any actually is (Aronson 2013). However, classical causality or, in the first place, a representation of quantum processes and developing a corresponding interpretation of quantum phenomena and QM is, again, difficult to maintain, in view of several experimentally well-established features of quantum phenomena. I shall mention two of them here: the uncertainty relations and quantum correlations, both independent of QM and pertaining to quantum phenomena themselves.

The uncertainty relations make it more difficult to define a classical-like state of the system considered, which appears to be required for causality, albeit, again, not altogether impossible. Thus, the uncertainty relations are fully valid in Bohmian mechanics, which is realist and causal, but, it follows, not deterministic (in its predictions coincide with those of QM). Bohmian mechanics is, however, mathematically different from QM and, in addition, expressly violates the requirement of locality. The latter dictates that physical systems could only be physically influenced by their immediate environment or that the instantaneous transmission of physical influences between spatially separated physical systems, found in Bohmian mechanics, is forbidden. I shall discuss locality later, merely noting for now that, although locality is a more general concept, a violation of locality implies a violation of relativity, the validity of which is well established experimentally, which is one of the reasons why a violation of locality, such as that found in Bohmian mechanics, is commonly (there are exceptions) seen as undesirable. The requirement of locality adds to the difficulties of establishing a causal or, in the first place, realist theory of quantum phenomena in view of quantum correlations, such as those of the EPR type, which combine indeterminacy or randomness of individual quantum events and correlations found between these events, when the corresponding experiment is repeated multiple times. Interpretations of quantum phenomena and QM in the spirit of Copenhagen avoid these difficulties, in part, by being local alternatives to realism, because realism appears (the question is under debate, along with that of locality of quantum phenomena and QM) to imply nonlocality in view of the Bell and Kochen-Specker theorems and related findings, on which I comment below.

The *structural* nature of the nonrealism adopted here is defined by the fact that the effects of quantum objects and processes (which cannot be represented or even conceived and, hence, cannot be assigned structure) on the world we experience, specifically via measuring instruments affected by their interactions with quantum objects, have structure in the sense of the combination of indeterminacy or randomness and a correlational statistical order. It may be helpful to briefly consider both the difference (which is irreducible) between, and yet the common foundation of, "structural *nonrealism*" and the philosophical trend known as "structural *realism*," a reference that is difficult to avoid once the heading structural nonrealism is used. Roughly, structural realism states, first, in the epistemic (more Kantian) version, that all we know is the structure of the relations and not the things themselves, and, second, in the more radical ontological version, that there are no "things" and that structure is all there is (Ladyman 2016). Heisenberg, in his later writings, speaks of "the structure of matter" and of Kant's thing-in-itself as "a mathematical structure," a concept that is in accord with to structural realism, to which Heisenberg appears to be closer at this point of his thinking, as against the time of his creation of quantum mechanics in 1925 (Heisenberg 1962, pp. 147-166, 91).

Structural *nonrealism* is based, just as is structural realism, on the concept of *reality* (essentially referring to that which exists), but, in contrast to structural realism, it conceives of this reality as a "reality without realism," RWR, thus defining the corresponding, RWR, overall epistemological view and the corresponding interpretation of QM or higher-level quantum theories. By this I mean that one cannot know or even conceive, now or possibly ever, of the ultimate character of this reality, either the reality of "things" or the reality of the relationships between "things," ans thus structures, defining this reality. Although, as will be seen, a certain *sense* of "thing-ness" may apply in the case of "quantum objects," or more accurately, what is idealized as such, *no concept* of thing or object, or relationships between things is applicable to this reality, any more than any other concept, ultimately those of reality and existence included. Accordingly, one can only speak of quantum objects or their thing-ness as an idealization, an idealization that is more radical than Kant's conception of noumena or things-in-themselves, which, while unknowable, are still assumed to be thinkable (Kant 1997, p. 115).

It follows that a "reality without realism," which defines the ultimate (quantum) reality, is a reality without structure, as the concept of structure is commonly understood or, for that matter, in any possible understanding of this concept. Nevertheless, the *nonrealism* to be considered in this article is *structural*. First, it depends on mathematical structures, in the conventional sense, found in and defining the mathematical models (a concept close to that of mathematical structure), which do not have a representational role of the kind they do in structural realism. Secondly, and most crucially (because it would make this nonrealism structural even as an interpretation of quantum phenomena themselves apart from quantum theory), while it cannot apply to the ultimate reality at stake, the concept of structure in the extended sense of this article does apply at the level of quantum phenomena, by virtue of considering formations, quantum correlations, that combine indeterminacy or randomness with organization and thus structure, albeit only of a statistically correlational character. The mathematical structure of the mathematical formalism of QM relates to these correlational structures by means of correctly predicting them. It follows, however, that the ultimate workings of nature responsible for the emergence of these correlational structures are, as against structural realism, beyond our knowledge or even conception. In other words, while there is no way to assign properties, structure, or laws to the efficacity of these correlational effects, these effects themselves are organized, possess structure, compelling this article to extend the term structure to designate the character of quantum information, which combines organization and indeterminacy or randomness.

Structural *nonrealism*, thus, also contains, as its part, a form of *realism* (pertaining to both things and the relationships between things), applicable at the level of effects or phenomena, but, as against structural realism, not at the level of the efficacity of these effects. Although it might appear that epistemic structural realism meets, via Kant, structural nonrealism halfway, this is not the case, because the reality at stake in structural nonrealism is more unreachable than that of Kant's noumena or things-in-themselves. Besides, as just noted, while structural nonrealism allows for a certain *reality* and in this sense thingness at the ultimate level of reality, it does not allow for a realist representation or even specified conception of either this thingness and its constitution or the relationships between any parts of this constitution. The term realism customarily refers to the possibility of a representation or at least conception of the reality considered, a reality of things or a reality of structures. Structural nonrealism denies either possibility at the ultimate level, and only allows for the realism of things and relationships between them, and thus structures, at more surface levels, such as those of structure at the level of quantum phenomena that define the architecture of quantum information, the architecture that is quantum information, as a specific form of organization of classical information. The existence of quantum objects, or, again, the necessity of idealization of that which exists thus designated, is inferred from this nonclassical architecture. They are "it from bit," famously invoked by Wheeler in his visionary manifesto of quantum information theory, in the present view, the unknowable or even unthinkable "it," inferred from the knowable "bit," from the structures of bits of quantum information obtainable from measuring instruments (Wheeler 1990, p. 3).

It follows that structural nonrealism is fundamentally, irreducibly different from structural realism of either type, epistemic or ontic. What provides the common foundations underlying both philosophical positions, even in their irreducible difference, is the concept of structure, apparently unavoidable in modern physics, from Galileo on, classical, relativistic, or quantum, and one might surmise, whatever may come beyond them. I would argue, however, that this irreducibility of structure in our interactions with nature, or in our thought in general, is defined by *human nature*, by our capacity for what we call thought, conscious and unconscious, rather than, as in (at least ontic) structural realism, belonging to nature apart from us. By this, given that our nature is still nature, I mean the nature that we assume existed before we existed and will exist when we no longer exist. This assumption will serve here as the definition of dead matter, the ultimate constitution of which cannot, in the present, RWR-type, view, be assigned structure.

This dead matter, at some point of its history, gave rise to living matter and then to us, and still keeps us alive, probably not for very long on the scale of its history, from the Big Bang on. Life, which is much older than we are, is likely to last longer as well. For now, however, life, as manifested in the neurological functioning of our brains, enables us to *think* the nature and structure of quantum phenomena and to create mathematical structures that allows us to predict the data observed in them. It also enables us, in interpreting both structures, to infer the *unthinkable* nature, thus no longer structure, of that (designated as quantum objects) which gives rise to these phenomena, or to interpret this nature and this structure otherwise. Thus, it enables both structural nonrealism and structural realism, equally dependent on thinking in terms of structures, but irreducibly different in thinking how the structures at stake in quantum theory come about.

### 3. Physics and Philosophy of Structural Nonrealism: From Realism to Reality without Realism

This section outlines the key concepts of structural nonrealism. Given that the relationships among these concepts affect how they work or even how they are ultimately defined, the initial definitions for some of them are bound to be provisional. A fuller sense of these concepts emerges in their relationships to other concepts as my discussion proceeds, even beyond this section. I understand a physical theory as an organized assemblage of physical, mathematical, and philosophical concepts, associated with physical objects or phenomena.

It is fitting to begin with the concept of concept, rarely adequately considered in physical or even philosophical literature. The present definition or concept of concept follows G. Deleuze and F. Guattari (Deleuze and Guattari 1994). In this definition, a concept is not merely a generalization from particulars (which is commonly assumed to define concepts) or a general or abstract idea, although a concept may contain such ideas, specifically abstract mathematical ideas in physics, or in mathematics itself, where such ideas may become concepts in the present sense. (By an "idea" I refer to any mental construct, keeping in mind that there are more refined definitions, including those closer to that of concept here.) A concept is a multicomponent entity, defined by the *organization* of its components, which may be general or particular, and some of these components may be concepts in turn. A concept is, thus, a *structure*, and it is the relational organization of its components that defines it. This is yet another structural dimension of structural nonrealism, as always conceptual, and yet another reflection of the necessary role of structure in our thinking.

Consider the concept of "tree," even as it is used in our daily life. On the one hand, it is a single generalization of all (or most) particular trees. On the other hand, what makes this concept that of "tree" is the implied presence of further elements, components, or sub-concepts, such as "branch," "root," "leaf," and so forth, and the organized relationships among them, which gives it structure. The concept of tree acquires further features and components, indeed becomes a different concept, in philosophy or in science, such as botany and biology.

This type of change is characteristic of the scientific use of concepts or terms derived from the concepts of daily life but defined otherwise, for example in quantum theory, even apart from obvious cases such as "charm" or "color," used more or less whimsically and thus without causing any confusion, although they still exemplify this point. More interesting are cases when a confusion is possible and even common, as in considering Bohr's concepts. As Bohr noted, in referring to his interpretation, "words like 'phenomena' and 'observations,' just as 'attributes' and 'measurements,' are used [in his interpretation] in a way hardly compatible with common language and practical definition," in contrast to classical physics, which, while refining, specifically mathematically, our daily concepts, such as "motion," still retains this compatibility (Bohr 1987, v. 2, pp. 63–64). This compatibility is retained when these concepts are applied, as they are in Bohr or here, only to the observed parts of measuring instruments, which observed parts are described by classical physics. But, which is often missed in considering Bohr's argumentation, it is no longer retained in considering quantum objects, including those parts of measuring instruments through which the latter interact with quantum objects and which are assumed to be quantum. The same situation obtains in the case of complementarity, a word that does not appear to have been used as *a noun* before Bohr (as opposed to the adjective "complementary") and was introduced by Bohr to designate a new concept. As he said: "In the last resort an artificial word like 'complementarity' which does not belong to our daily concepts serves only briefly to remind us of the epistemological situation here encountered, which at least in physics is of an entirely novel character" (Bohr 1937, p. 87). In other words, complementarity, which I shall properly define below, is a new physical concept with several interrelated components. As most innovative scientific concepts, complementarity, when introduced, was not defined by generalization from available entities: it was something entirely new. It then functioned, in part, by generalizing multiple specific entities, such as specific complementary configurations, say, those of the position or the momentum measurement, always mutually exclusive at any given moment of time in the case of quantum phenomena. While this mutual exclusivity is an essential feature of complementarity, Bohr's concept is, as will be seen, more complex and multicomponent, and cannot be limited to this feature.

Simple (single-component) concepts are rare, if possible at all, as opposed to appearing as such because their multicomponent structure is provisionally cut off, as in the case of the concept of tree, as just noted. In practice, there is always a cut off in delineating a concept, which results from assuming some of its components to be primitive entities, whose structure is not specified. These primitive components could, however, be specified by a further delineation, which could lead to a new overall concept, containing a new set of primitive (unspecified) components. Such changes may result from critical reexaminations of such primitive, uncritically used, components or subconcepts. The history of a concept is a history of such successive specifications, and every concept, however innovative, has a history.

The same type of process defines the history of a given theory, and just as every concept, every theory, however innovative, has a history, from which it emerges. By a theory I understand an organized assemblage of concepts, as just defined, under the assumption that a theory could be modified in the course of its history by modifying its concepts or the relationships among them. A viable physical theory must, however, relate to a given multiplicity of phenomena or objects, which are customarily assumed to form the reality considered by this theory.

I refer to both phenomena and objects, because, as Kant realized, they are not the same even in classical mechanics, which deals with individual classical objects or sufficiently small classical systems. However, as Kant realized as well, when a direct observation is possible, classical objects, say, planets moving around the Sun, and our phenomenal representation of them could be treated as the same for all practical purposes, because our observational interference with their behavior could, at least in principle, be neglected or compensated for, thus allowing us, in principle, to consider this behavior independently. Indeed, this was assumed to be possible in the case of all classical physical objects, even when they were not or even could not be observed, as in the case of atoms or molecules in the kinetic theory of gases in the nineteenth century. Quantum phenomena put this assumption into question. Quantum phenomena themselves, defined by the effects of the interactions between quantum objects and measuring instruments, are observable in the same way as are classical physical objects and could be treated as such, as *classical objects*. By contrast, the "uncontrollable" nature of these interactions precludes any observation or, at least in nonrealist, RWR-type, interpretations, even rigorous inferential reconstitution of the independent nature and space-time behavior of quantum objects (Bohr 1935, pp. 697, 700). Such a reconstitution is, in Bohr's words, "*in principle* excluded" (Bohr 1987, v. 2, p. 62). Nobody has ever observed, at least thus far, an electron or photon as such, in motion or at rest, to the degree that such a concept, as opposed to a change of a state of an electron or photon, ultimately applies to them, or any quantum objects, qua quantum objects, no matter how large. (Photons, of course, only exist in motion.) It is only possible to observe traces, such as spots on photographic plates, left by their interactions with measuring instruments. The present interpretation, in which quantum objects and processes are beyond conception (rather than only representation), makes this difference more radical than that defined by Kant, for whom objects, as noumena or things-in-themselves, are only beyond knowledge and representation but not beyond conception (Kant 1997, p. 115). As will be seen below, Kant gives to the faculty he calls Reason [*Vernunft*] the power to claim that its thinking concerning noumena or things-in-themselves corresponds to what is in fact true, although some of such truths may not actually represent, at least not completely, the nature of these truths.

The history of a theory is accompanied and shaped by the history of its interpretations, which are essential (if sometimes implicit) for establishing the relationships between a theory and the phenomena or objects it considers. The history of QM, in particular, has been shaped by a seemingly uncontainable proliferation of, sometimes conflicting, interpretations.[7] In fact, a theory always involves, however implicitly, an interpretation, for example and, in modern physics, in particular, in defining how the mathematics used by a theory works in establishing these relationships, which is customarily done by means of mathematical models.

---

[7] It is not possible to survey these interpretations here. Each rubric on by now a long list (e.g., the Copenhagen, the many-worlds, consistent-histories, modal, relational, transcendental-pragmatist, and so forth) contains different versions. The literature dealing with each interpretation is immense.

I define a *mathematical* model as a mathematical structure or a set of mathematical structures that enables such relationships.[8] If a mathematical model consists of a single mathematical structure, this model is pretty much equivalent to this structure, as reflected in the concept of model in mathematics itself, roughly, a collection, usually, a set, of abstract mathematical objects and relationships among them that satisfy given axioms and rules of procedure. In physics, these relations between a model and the objects or phenomena considered may be representational, which makes them realist, and derive their predictive capacity from their representational nature, as in the case of models used in classical mechanics or relativity; or they may be strictly predictive, without being representational, as in QM in nonrealist interpretations, where our predictions are, in general, probabilistic or statistical in character. As just noted, a theory always involves an interpretation of itself and of the model or models it uses, in the latter case especially by establishing the way in which these models relate to the phenomena or objects considered. Rigorously, a different interpretation of a given mathematical model defines a different theory, in the present definition of theory. For simplicity, however, I shall speak of the corresponding interpretation of the theory itself, containing a given mathematical model, interpreted by this theory, say, of an interpretation of QM, in considering an interpretation of its mathematical formalism.

I shall now define a concept of reality, which is very general, and, as such, equally applies to the concepts of reality found in both structural realism and structural nonrealism. Indeed, it arguably applies to most, even if not all (which would be impossible), currently available versions of realism and nonrealism, scientific or philosophical. Each version of either would give this concept further features, most especially, by assuming this reality to be representable or at least conceivable in realism and as being beyond representation, or as here, even beyond conception, in nonrealism. By *reality* I refer to that which exists or is assumed to exist, *without making any claim concerning the character of this existence*, the

---

Standard reference sources, such as *Wikipedia* ("Interpretations of Quantum Mechanics") lists most common rubrics.

[8] As other major concepts discussed here, the concept of a mathematical model has a long history, which is also a history of diverse and often diverging definitions, and literature on the subject is extensive. It is not my aim to discuss the subject as such or engage with this literature, which would be difficult within the scope of this article. It is also not necessary. The present concept of a model, while relatively open, is internally consistent and is sufficient to accommodate those models that I shall consider. See (Frigg and Hartmann 2012), which addresses the subject on lines of analytic philosophy, and (Plotnitsky 2016), which offers a more comprehensive analysis of the concept of mathematical model used here.

type of claim that defines realism. I understand existence as a capacity to have effects on the world with which we interact and that, because it exists, has such effects upon itself, effects that need not have classical causes.

In physics, the primary reality considered is that of nature or matter, including that of fields or that to which the concept of field, classical or quantum, would relate, keeping in mind that the concept of quantum field in QFT is, beginning with its definition, a subject of much debate and controversy.[9] This or any other idea of nature or matter is still a product of thought, the reality of thought, which reality, however, is customarily assumed to be a product of the material processes in the brain, and thus of matter. "Matter" is, generally, a narrower concept than "nature," although when at stake is, as for the most part in this article, the ultimate level of the *material* constitution of nature, both concepts merge. Matter is commonly, but not always (although exceptions are rare), assumed to exist independently of our interaction with it, and to have existed when we did not exist and to continue to exist when we will no longer exist, an assumption that may, as noted, be seen as defining the independent existence of matter.

This assumption is upheld in nonrealist interpretations of QM and in structural nonrealism, but in the absence of a representation or even conception of the character of this existence, for example, as either discrete or continuous. (It is not possible to conceive of an *elemental* entity that is both: a composite entity can have both elements.) In such interpretations, discreteness only pertains to quantum phenomena, observed in measuring instruments, and continuity has no physical significance at all. It is only found in the formalism of QM, which, while continuous in its mathematical character, relates to discrete quantum phenomena by predicting, with the help of rules, such as Born's rule, added to this formalism, the probabilities or statistics of the occurrence of quantum events.

Physical theories prior to quantum theory have been, and have been designed to be, realist theories, usually *representational* realist theories. Such theories aim to represent, usually (classically) causally, the corresponding objects and their behavior by mathematical models, assumed to idealize how nature works, an assumption sometimes referred to as "scientific realism," commonly under a broader assumption that a scientific theory or a set of scientific theories can, at least ideally, represent nature, hopefully also in its ultimate constitution. Thus, classical mechanics (used in dealing with elemental individual objects and some small classical systems), classical statistical mechanics (used in dealing, statistically, with large classical systems, whose individual constituents are assumed to be described by classical mechanics), or chaos theory (used in dealing with classical systems that exhibit a highly nonlinear behavior) are all realist theories. I shall be primarily concerned, apart from quantum theory, with these types of realist theories and the corresponding models.[10] In its broad philosophical, realism, representational or nonrepresentational (defined below), need not be scientific and, specifically, mathematical. It could also, for example, in certain biological theories, be scientific without being mathematical, but not without being conceptual.

---

[9] I considered QFT in detail in (Plotnitsky 2016), where quantum fields are defined in terms of "reality without realism," although not within the framework of structural nonrealism.

This irreducibility of conceptual mediation in scientific thinking, especially thinking concerning the nature of reality, was acutely realized by Einstein. He saw the practice of theoretical physics as that of the invention of new concepts, through which and, he insisted, only through which one can approach reality. He argued that a viable realist representation of physical reality could only be achieved by means of conceptual construction, rather than by means of observable facts themselves, a view that he saw as the empiricist "philosophical prejudice," exemplified by E. Mach's positivist philosophy. He added: "Such a misconception is possible because one does not easily become aware of the free choice of such concepts, which, through success and long usage, appear to be immediately connected with the empirical material" (Einstein 1949, p. 47). As Einstein was undoubtedly aware, such a choice is never entirely free, which is to say, is never entirely a choice. It might be more precise to speak of one's *decision* concerning which concepts to adopt. However, especially when to comes to major discoveries, such a decision may involve enough freedom to justify Einstein's claim. On the other hand, he is entirely correct in arguing for the necessity of concepts in order to understand construct the structure of reality, even if, as in nonrealist, RWR-type, views, no such structure could be assigned to reality. One still needs concepts, such as that of structure to rigorously establish this impossibility, as an inference from the structure of information observed in measuring devices, impacted by quantum objects, or, again, what is so inferred, as defining this reality beyond representation or conception, including that of structure.

In his *Physics and Philosophy*, an important book invoked at several junctures in this article, Heisenberg used this type of emphasis on the role of concepts in theoretical physics to argue that "the Copenhagen interpretation of quantum theory" was not positivistic, in which he followed Einstein, with whom he had important exchanges concerning the subject and the relationships between observation and theory in physics, following Heisenberg's invention of quantum mechanics (Heisenberg 1962, pp. 45-46; Heisenberg 1989, p. 30; Plotnitsky 2016, pp. 42-44). The phrase "the elements of reality" used by Heisenberg in the passage I am about to cite is, too, borrowed from Einstein, who often used it, most famously in the EPR paper (Einstein et al 1935, p. 138). Heisenberg says: "The Copenhagen interpretation of quantum theory is in no way positivistic. For, whereas positivism is based on the sensual perceptions of the observer as the elements of reality, the Copenhagen interpretation regards things and processes which are describable in terms of classical concepts, i.e., the actual, as the foundations of any physics interpretation" (Heisenberg 1962, p. 145). "The actual" here refers, following Bohr, to what is observed in measuring instruments. I shall return to the role (often misunderstood) of classical concepts in Bohr's argumentation later. My main point at the moment is the fundamental role of concepts in theoretical physics. Besides, QM involved plenty of concepts, physical, such as complementarity, and mathematical, such as noncommuting operator variables, that are not classical. While in the spirit of Copenhagen, "the Copenhagen interpretation" invoked here and discussed in Heisenberg's book is shaped by a mixture of Heisenberg's own and Bohr's views, from which Heisenberg departs at certain points, in particular, in his understanding of complementarity and in introducing the concept of "potentiality" or "*potentia*," not found in Bohr. Heisenberg, at this stage, also appears to have inclined to believe in the capacity of mathematics to represent the ultimate structure of matter, a belief not shared by Bohr either. Bohr would, however, have agreed that his understanding of the quantum-mechanical situation was "in no way positivistic," in part for the reasons stated by Heisenberg.

---

[10] Their status as realist could of course be questioned, specifically on Kantian lines. For example, to what degree do the spacetimes of relativity correspond to the ultimate constitution of nature? How realist is such a representation, even as an idealization, or in the first place, the concept of space-time itself, as opposed to serving as a mathematical tool for correct predictions, in this case, exact rather than probabilistic, as against quantum theory? These questions could also be posed in the case of special relativity, or even classical mechanics, in which case, as indicated earlier, the representational idealizations used are more in accord with our phenomenal experience, which is only partially the case in relativity. For some of these complexities in relativity, see (Butterfield and Isham 2001). However, these cases still allow for viable idealized realist and causal, indeed deterministic, models. This appears to be much more difficult, if possible at all, in quantum theory (QM, QFT, or QTFD), which, at least thus far, has to be irreducibly probabilistic on experimental grounds even in considering elemental individual quantum objects and processes. This excludes a deterministic model, although it does not exclude a causal or, in the first place, realist model.

Indeed, Bohr's interpretation was in fact concerned, even primarily concerned with quantum objects, even though they could not be observed as such, but only inferred from their effects on measuring instruments. It is not us but nature that, in its interaction with us, via our experimental technology (which includes our bodies that observe these effects but which of course consists of so much more) that is responsible for these effects, even though we cannot know or possibly even conceive how these effects come about. It is true that we, our bodies, and out technology are nature, too. But neither are sufficient in themselves to produce these effects, which require the quantum constitution of nature apart from us and our technology, or, again, that which compels us to speak of this constitution, assuming that even the term "constitution" applies.

One could also define another type of realism, which is not representational. This realism encompasses theories that would presuppose an independent structure of reality governing the behavior of the ultimate objects these theories consider, while allowing that this architecture cannot be represented, even ideally, either at a given moment in history or perhaps ever, but if so, only due to practical limitations. In the first eventuality, a theory that is merely predictive may be accepted for lack of a realist alternative, but under the assumption or with the hope that a future theory will do better, in particular by virtue of being a realist theory of the representational type. Einstein adopted this view in considering QM, which he expected to be eventually replaced by a (representational) realist theory, ideally, a field theory of a classical-like type. Einstein followed this program, first, very successfully, in general relativity and then, without much success, in trying to unify gravity and electromagnetism. Even in the second eventuality, however, reality is customarily conceived on realist models of classical physics, possibly adjusting them to accommodate new phenomena and new concepts, as happened in the case of electromagnetism and the (classical) concept of field in the nineteenth century.

What unites both types of realism and thus defines realism more generally is the assumption that the ultimate constitution of nature possesses properties, beginning with that of the thingness of things, and the structured relationships among them, or (as in structural realism) just structures, that may be either (a) known in one degree or another and, hence, represented, at least ideally, by a theory or model, or (b) unknown or even unknowable. The difference between (a) and (b) is that between the two types of realism just described, representational and nonrepresentational.

The assumption of realism of either type is abandoned or even precluded in nonrealist, RWR-type, interpretations of quantum phenomena and QM, beginning with that of Bohr. In such interpretations, the mathematical model of QM, defined by its mathematical formalism, becomes a strictly probabilistically or statistically predictive model, while suspending or even precluding a representation and possibly a conception of quantum-level reality and (in the second eventuality, as an immediate consequence) an assumption that this reality is causal. Although realist interpretations of the quantum-mechanical formalism have been offered, its representational capacities and the effectiveness of such interpretations, or (it would be surprising otherwise) conversely, the effectiveness of nonrealist interpretations, have been and continue to be intensely debated.[11] The probabilistic or statistical character of quantum predictions must, however, be maintained by realist interpretations of these theories or alternative theories (such as Bohmian mechanics), to accord with what is observed in quantum experiments, where only probabilistic or statistical predictions are possible. This is because the repetition of identically prepared experiments in general leads to different outcomes, and, unlike in classical physics, this difference cannot be diminished beyond the limit defined by Planck's constant, $h$, by improving the capacity of our measuring instruments, as manifested in the uncertainty relations, which would apply even if we had perfect instruments.

---

[11] While primarily concerned with nonrealism, the present argument does not aim to deny realism, still a generally preferred philosophical view. One must also be mindful of additional complexities. One might, for example, argue against a realist interpretation of the wave function, as a continuous entity, and maintain a realist view of QM, as referring to a discrete ultimate ontology (Rovelli 2016). In the present view, no ontology, either continuous or discontinuous, can again be assigned at the ultimate level of reality. One could assign ontology, ultimately discrete, at the level of quantum phenomena, defined by what is observed in measuring instruments, and probabilistically predicted by means of QM, including the wave function.

Nonrealist interpretations do, again, assume the concept of *reality*, as that which is assumed to exist, without, in contrast to realist theories, making any claims concerning the *character* of this existence, which, in terms adopted here, makes this concept of *reality* that of "reality *without* realism" (Plotnitsky 2016, Plotnitsky and Khrennikov 2015).[12] Physically, an interpretation of this kind is an interpretation of quantum phenomena and QM, corresponding to a certain overall view in physics and philosophy—the RWR interpretation and the RWR view. Such an interpretation places quantum objects and processes beyond representation, which I shall term "the weak RWR view," or beyond conception, beyond the reach of thought altogether, which I shall term "the strong RWR view," adopted here. (When either form of the RWR view applies, I shall just refer to the RWR view, the RWR-type interpretation, and so forth.) The existence of quantum objects or something that leads to this idealization is inferred from the totality of effects they have on our world, specifically on experimental technology.[13] Nothing, however, could be said, or, again, in the strong RWR view, even thought, concerning *what* happens between quantum experiments.

As Heisenberg observed, it may be said that *something* does "happen," say, in an "atom," between our observations, as manifested in changes we observed in our instruments, and thus independently of us, but only insofar as we keep in mind the provisional nature of such words or concepts as "happen" or "atom," which are ultimately inapplicable in this case, any more than any other human words or concepts. Besides, that such a change was due to the interaction with the same quantum object cannot be guaranteed in any single experiments; it can only be ascertained statistically when the same experiments is repeated many times. According to Heisenberg:

> There is no description of what happens to the system between the initial observation and the next measurement. …The demand to "describe what happens" in the quantum-theoretical process between two successive observations is a contradiction in adjecto, since the word "describe" [or "represent"] refers to the use of classical concepts, while these

---

[12] One could, in principle, see the claim concerning merely the existence or reality of something to which a theory can relate without representing or even conceiving of it as a form of realism. This use of the term realism is sometimes found in advocating interpretations of QM that are nonrealist in the present sense. However, placing "reality" outside "realism" is advantageous, especially in the context of this article.

[13] For an analysis of the concept of quantum objects from a realist perspective, see (Jaeger 2015).

concepts cannot be applied in the space between the observations; they can only be applied at the points of observation. (Heisenberg 1962, pp. 47, 145)

The same, it follows, must apply to the word "happen" or any word we use, and we must use words and concepts associated to them, even when we try to restrict ourselves to mathematics as much as possible. There can be no physics without language, but quantum physics imposes new limitations on using it. Heisenberg adds later in the book: "But the problem of language is really serious. We wish to speak in some way about the structure of the atoms and not only about 'facts'—the latter being, for instance, the black spots on a photographic plate or the water droplets in a cloud chamber. But we cannot speak about the atoms in ordinary language" (Heisenberg 1962, pp. 178-179). Nor, by the same token, can we use, in referring to the atoms, ordinary concepts, from which our language is not dissociable, or for that matter any concepts, physical or philosophical.

On the other hand, as Heisenberg noted already in his Chicago lectures of 1929, mathematics, and one might add, for the reasons explained below, algebra in particular, is, "fortunately," free from the limitations of language, fortunately because, as Heisenberg stressed there, otherwise quantum mechanics would not be possible (Heisenberg 1930, p. 11). Mathematics also allows to circumvent the limits our phenomenal, representational intuition, also involving visualization, sometimes used, including by Bohr, to translate the German word for intuition, *Anschaulichkeit*. "Visualization" and its avatars, such as "pictorial visualization," is often invoked by Bohr, by way of this translation, in considering quantum objects and behavior, as being beyond our capacity to phenomenally represent them (e.g., Bohr 1987, v.1, pp. 51, 98-100, 108, v. 2, p. 59). It clear, however, that he sees quantum objects and processes are being beyond any representation, even if not, as here, conception, including, again, in contrast to Heisenberg, a mathematical one. In any event, in Bohr's and Heisenberg's view alike (on this point Heisenberg was in agreement with and in fact followed Bohr [Heisenberg 1930, p. 11]), our phenomenal intuition and, thus, concepts that it can actually form in physics, appear to be classical and, quite possibly, when dealing with change, classically causal. For, according to L. Wittgenstein, we might not be able to conceive of a process that is not causal (Wittgenstein 1924, p. 175). Certain concepts, such as complementarity, can help us to cope with these difficulties, but they will not enable us to entirely avoid them. It is mathematics that helps us most.

As free from these limitations of language and ordinary, or even philosophical intuition, mathematics could, in principle, be assumed to represent the structure of the atoms, as Heisenberg eventually came to believe, certainly by the time of the statement just cited in *Physics and Philosophy*, which gives mathematics at least some capacity to do so, on lines close to those of structural realism. On the other hand, as is clear from his statements and correspondence, cited in the next section, at the time of his discovery of quantum mechanics, he appears to have seen mathematics' freedom from these limitations of language and ordinary, while crucial for quantum mechanics and even making its invention possible, in terms its probabilistically predictive rather than representational capacity. Bohr adopted this view following Heisenberg's discovery of quantum mechanics and never relinquished it. It is not exactly clear when this change in Heisenberg's view, progressively more noticeable in his publications, occurred. He might have started to move toward this view already by the time of his Chicago lectures and even around at the time of his discovery of the uncertainty relations in 1927, following his exchange with Einstein, mentioned above, concerning the relationships between observation and theory in physics (e.g., Heisenberg 1962, pp. 45-46; Heisenberg 1989, p. 30, Plotnitsky 2016, pp. 42-44). In his several recollection of this exchange and his thinking leading to his discovery of the uncertainty relations, Heisenberg referred to the following perspective, which, he said, was important in this discovery: "Instead of asking: How can one in the known mathematical scheme express a given experimental situation? the other question was put: Is it true, perhaps that only such experimental situations can arise in nature as can be *expressed* in the mathematical formalism" (Heisenberg 1962, pp. 44-45; Heisenberg 1989, p. 30; emphasis added). Apart from the role of this view in Heisenberg's discovery of the uncertainty relations (which is a separate topic), this assumption would imply that the mathematical formalism of quantum mechanics, represents, "expresses," at least ideally, the quantum workings of nature, rather than merely enables us to predict the effects of the interactions between quantum objects and measuring instruments, as Heisenberg thought at the time of his discovery of quantum mechanics.

It may be added in this connection, that the following view has been advanced, beginning with both Dirac and von Neumann (Dirac 1930; von Neumann 1932), which were influential in shaping and promoting this view, sometimes even characterized as "the Copenhagen interpretation." According to this view, the mathematical formalism of QM represents quantum processes between quantum experiments, moreover causally, with randomness and thus the recourse to probability only brought in by measurement. At least, this formalism would represent them mathematically, that is, in the absence of a physical conception of these processes, although, as noted, causality would entail a partial conception of them. In Heisenberg's view, again, at the time of his discovery of QM, or that of Bohr, *in most of his thinking concerning the subject*, the quantum-mechanical formalism does not represent anything, either between or during measurements; it only predicts, probabilistically or statistically, what could happen in certain possible experiments on the basis of the data obtained in certain other, previously performed, experiments.[14]

---

[14] I add the emphasized qualification, because Bohr briefly and still ambivalently entertained this type of view in his early thinking (in 1926-1927) concerning QM and complementarity, which may be one of the reasons why this view is sometimes presented as "the Copenhagen interpretation." Bohr, however, quickly abandoned this view, in part under an impact of his initial exchanges with

The present, strong, form of the RWR view is manifestly more radical than that of Kant's noumena or things-in-themselves, vis-à-vis phenomena or appearances formed in our minds. According to Kant, while noumena are unknowable, they are still in principle conceivable, especially by what he calls "Reason" [*Vernunft*], a higher faculty than "Understanding" [*Verstand*], which only concerns phenomena (Kant 1997, p. 115, Plotnitsky 2016, pp. 19-21). As noted above, the Kantian Reason is even able to form true conceptions of the ultimate nature of reality (of things-in-themselves). Kant's scheme is, thus, realist in the present definition: it conforms at least to realism of a nonrepresentational type, and in the case of Reason sometimes even of a representational type, although Reason's claims concerning the noumenal reality may not entail a specific representation or even a definite conception of it, which may be impossible. Thus, the claim of the existence of God is true, according to Kant, but, apart from viewing God as a supreme agency of creation, little, if anything, could be known or thought by us about the ultimate nature of this existence, which led Kant to his criticism of all claims to the effect (Kant 1997). Kant's God is the God of metaphysics, not the (personal) God of Religion (Badiou 2007, pp. 21-32).

It might indeed appear that the present (RWR) view of matter is similar to Kant's or related views of the divine. This is not the case, however, because in the present or most other RWR-type views, no conception of divine or divine like creation can be applied to matter (except metaphorically), any more than any other property. In this type of view, matter has effects, but, given the character of some of these effects, manifested in quantum phenomena, the ultimate efficacity of these effects cannot be represented in terms of any properties or even concepts, for example, that of some supreme being. It is clear that any such concept, even that of God as a mathematician, is not only classical but also anthropocentric, something that, as Bohr warned on many occasions, one should be cautious in applying to the ultimate constitution of nature. Einstein's famous "God does not play dice" would be one example here, except that Einstein clearly uses it metaphorically (meaning nature), a brilliant rhetorical move that immortalized the statement.[15]

It might be added that the weak RWR view, at least in Bohr's version of it, is still different from that of Kant. For, there is, in Bohr's view, no power, such as that of the Kantian Reason, that would allow us to have such a conception at the noumenal level (the

---

Einstein concerning the subject (Plotnitsky 2016, pp. 66-68, 124-131, 159). By contrast, as just indicated, Heisenberg appears to have started around the same time and also under an impact of his exchanges with Einstein's ideas, to move toward a certain mathematical form realism, eventually on the lines of structural realism. Einstein, while stressing the defining significance of mathematics in forming physical concepts, believed that our physical concepts should, as physical (rather than only mathematical) concepts, have power to approximate the ultimate nature of reality.

[15] For a further discussion of this point, see (Plotnitsky 2016, pp. 65-66).

level of things-in-themselves), independently of a theory dealing with the observed phenomena, in this case, quantum theory and quantum phenomena. This philosophical position was also adopted by Heisenberg, expressly against Kant (Heisenberg 1962, pp. 86-92). In the present, strong RWR, view, as *things stand now*, such as conception is, again, strictly precluded.

The strong RWR view is, then, beyond all realism (as defined here), by virtue of placing the ultimate nature of reality, such as that of quantum objects and processes, beyond conception, whether such a conception be philosophical, physical, or mathematical. As noted earlier, a mathematical conception may be divorced from any physical or philosophical conceptuality. This makes such words as "ultimate nature," or "objects," "process," and "quantum," or even "reality" or "existence," provisional and ultimately inapplicable as well, insofar as our thinking assigns, as it perhaps bound to do (at least unconsciously) any specific representational concepts to them. "Reality" functions as a name without a concept. It designates something that is beyond thought but that, nevertheless, has manifested effects upon the world with which we interact by means of our bodies and technologies. But then, our bodies could be seen as technologies as well, including those of perception and thought, but far from only them. The history of physics has always been dependent on the mechanical technologies of our bodies, and it still is, despite the nearly unimaginable complexities of our experimental technologies, such as that of the present-day accelerators, and the ubiquitous role digital technologies play in present-day experimental and theoretical physics. Our bodies are still there to do physics, which is not possible without them, unless one ascribes (and some do) the capacity of doing physics to computers or (this happens as well) to nature itself.

In the strong RWR view, the ultimate reality at stake is not merely something that, while as yet unthought, might eventually be thought, be conceived of in any form, for example, by assuming that it has a structure even without specifying this structure. Instead, this reality is seen as that which cannot ever be reached by thought. If a conception of this reality, in quantum theory, the physical or mathematical, of the nature of quantum objects and processes is assumed to be formed (as some claim is the case even now), one reverts to realism, at least, of a nonrepresentational type, or if this conception is mathematical (with or without a physical conception this mathematics represents) likely of a representational type. I shall further comment on this point below.

As indicated earlier, it is not clear whether Bohr or Heisenberg held the strong RWR or the weak for of the RWR view. While the weak RWR view is philosophically different, it still allows one to see quantum theory as nonrealist, even as against nonrepresentational realism, given that, as things stand now, no concept of the ultimate (quantum) reality considered can be unambiguously used in quantum theory. Bohr and Heisenberg (in his earlier thinking, influenced by Bohr) might, however, have assumed the strong RWR view, because of their emphasis on the lack of causality in quantum processes, if one agrees with Wittgenstein's contention, mentioned earlier, that we cannot conceive of processes that are not causal (Wittgenstein 1924, p. 175). It is, however, not clear either whether either Bohr or Heisenberg thought this to be the case.

As explained earlier, assuming this to be the case does not prevent the possibility that a mathematical model would represent quantum behavior, perhaps noncausally, because a mathematical conception can do so in the absence of a physical conception corresponding to this mathematical conception, thus, entailing a strictly mathematical ontology of the ultimate physical reality, even if its physical character may not be entirely known or is not known at all, which is not an uncommon view, found, for example, in structural realism or in Heisenberg's later thinking. A form of realism advocated by Heisenberg in his later works still precludes any form of representational language or concepts, apart from mathematical ones (Heisenberg 1962, pp. 145, 167-186). For instance, this reality may be claimed to be invariant under a symmetry group, although all of the specifics may not be fully known, a view, again, found in Heisenberg's later writings. In the present view, the symmetry groups of quantum theory are still seen only as part of the probabilistically predictive mechanism of quantum theory, although symmetries may, of course, apply representationally at macro levels. Bohr's position (or, again, that of Heisenberg at earlier stages of his thinking about quantum theory) was defined by the view that a representation or even an application of any available or even possible conception, philosophical, physical, or mathematical, to quantum level reality is, in his words, "*in principle* excluded," as things stands now (Bohr 1987, v. 2, p. 62).[16]

"As things stand now" is, again, a crucial qualification, equally applicable to the strong RWR view, even though it might appear otherwise, because this view precludes any conception of reality not only now but also ever, by placing it beyond thought altogether. Nevertheless, the qualification "as things stand now" still applies. This is because a return to realism is possible, either on experimental or theoretical grounds. Thus, quantum theory, as currently constituted (QM, QFT, and QTFD), may be replaced by an alternative theory that allows for or requires a realist interpretation, or the RWR-type interpretations, either of the weak or the strong RWR type, may become obsolete, with quantum theory in place in its present form. There is, however, a difference in how this change would affect the weak vs. the strong RWR view, which is as follows. In the case of the weak RWR view, the mathematics of quantum theory would be brought into an unambiguous agreement with one or another conception of reality, possibly by way of representing this reality, on the line of Einstein's ideal of a fundamental physical theory and fulfilling his hope. In the case of the strong RWR view, the very *conception of reality beyond thought* would no longer be applicable to quantum theory. (This conception may still be entertained philosophically or apply elsewhere.) As things stand now, however, either view is interpretively possible, without affecting the physics or mathematics involved; and it is also possible that the RWR view, in either weak or strong version, will remain part of our future fundamental theories, as the development of QFT theory might indicate. QFT has been open to the RWR-type view and the corresponding interpretation from its inception with Dirac until now and was used, specifically by Bohr, in support of this view (e.g., Bohr 1987, v. 2, p. 64; Plotnitsky 2016, pp. 207-246). The RWR view, in either version, does not preclude the development of new fundamental theories, for example, that of quantum gravity. It only changes the nature of these theories and, in this first place, the nature of our thinking in physics and beyond by making the unrepresentable and, in the strong RWR view, the unthinkable an irreducible part of thought.

---

[16] As noted above (note 14), Bohr briefly entertained the idea that the ultimate nature of quantum reality could be represented by the mathematical formalism of QM, and moreover causally, in which case, it would be at least partially conceivable as well. Bohr, again, quickly abandoned the idea.

I now turn to the question of causality. As noted, RWR-type interpretations make the absence of classical causality nearly automatic. This absence is strictly automatic if one adopts the strong RWR view, which places the ultimate nature of reality beyond conception, because the assumption that the ultimate nature of reality is classically causal would imply at least a partial conception of this reality. However, even if one adopts the weak RWR view, which only precludes a representation of this reality, classical causality is still difficult to maintain in considering quantum phenomena. This is because to do so one requires a degree of representation, analogous to that found in classical physics, that appears to be prevented, in particular, by the uncertainty relations (which are independent of QM, although fully consistent with it). Schrödinger aptly expressed this difficulty, while disparaging the spirit of Copenhagen as "the doctrine born of distress," in his cat-paradox paper: "if a classical state does not exist at any moment, it can hardly change causally," where a classical state is defined by the ideally definite position and the definite momentum of an object at any moment of time (Schrödinger 1935a, p. 154). According to Bohr (who here refers by causality to what I call "classical causality"):

> The unrestricted applicability of the causal mode of description to physical phenomena has hardly been seriously questioned until Planck's discovery of the quantum of action, which disclosed a novel feature of atomicity in the laws of nature supplementing in such unsuspected manner the old doctrine of the limited divisibility of matter. Before this discovery statistical methods were of course extensively used in atomic theory but merely as a practical means of dealing with the complicated mechanical problems met with in the attempt at tracing the ordinary properties of matter back to the behaviour of assemblies of immense numbers of atoms. It is true that the very formulation of the laws of thermodynamics involves an essential renunciation of the complete mechanical description of such assemblies and thereby exhibits a certain formal resemblance with typical problems of quantum theory. So far there was, however, no question of any limitation in the possibility of carrying out in principle such a complete description…. Due to the essentially statistical character of the thermodynamical problems which led to the discovery of the quantum of action, it was also not to begin with realized, that the insufficiency of the laws of classical mechanics and electrodynamics in dealing with atomic problems, disclosed by this discovery, implies a shortcoming of the causality ideal itself. (Bohr 1938, pp. 94–95)

Elsewhere, Bohr notes: "[E]ven in the great epoch of critical [i.e., post-Kantian] philosophy in the former century, there was only a question to what extent a priori arguments could be given for the adequacy of space-time coordination and causal connection of experience, but never a question of rational generalizations [such as complementarity] or inherent limitations of such categories of human thinking" (Bohr 1987, v. 2, p. 65). By contrast, as he argues in the same article, "Discussion with Einstein on Epistemological Problems in Atomic Physics": "it is most important to realize that the recourse to probability laws under such circumstances is essentially different in aim from the familiar application of statistical considerations as practical means of accounting for the properties of mechanical systems of great structural complexity. In fact, in quantum physics we are presented not with intricacies of this kind, but with the inability of the classical frame of concepts to comprise the peculiar feature[s] of the elementary [quantum] processes" (Bohr 1987, v. 2, p. 34). While "the classical frame of concepts" may refer here to those of classical physics, Bohr may also be here closer to the strong RWR view. For, at this stage in his thinking about quantum theory (in 1940s), he argues that all concepts ("object," "process," and "quantum," among them) we can form are essentially classical and, as such, representational and, as noted above, via Wittgenstein, possibly even causal in nature, by virtue of our inability to conceive a process that is not causal (Wittgenstein 1924, p. 175). It is also in this article that Bohr, in the statement partly cited earlier, responds to Einstein's discontent with the nonrealist epistemology of QM and its irreducibly probabilistic or statistical nature. Einstein appears to have in mind something like the weak RWR interpretation, to which, in his view, QM lent itself, while that was, for Einstein, not necessarily the case for quantum phenomena themselves, which he saw as, in principle, open to a more classical treatment. Bohr says: "Even if such an attitude [that of Einstein] might seem balanced in itself, it nevertheless implies a *rejection* of the whole argumentation exposed in the preceding [essentially Bohr's nonrealist interpretation, now closer to the present, strong, RWR view], that, in quantum mechanics, we are not dealing with an arbitrary renunciation of a more detailed analysis of atomic phenomena, but with a recognition that such an analysis is *in principle* excluded" (Bohr 1987, v. 2, p. 62; emphasis on "rejection" added). This is a rejection, a rejection based on a philosophical view and not a counterargument. Einstein admitted as much because he saw Bohr's argumentation or the situation it reflected "as *logically possible without contradiction*, but …so contrary to [his] scientific instinct that [he could not] forego the search for a more complete conception" (Einstein 1936, p. 349; Bohr 1987, v. 2, p. 62; emphasis added). By a more complete conception he clearly meant a realist theory that would also avoid probability in considering elementary individual processes and events.

Heisenberg, too, argued that "the statistical nature of the laws of microscopic [quantum] physics cannot be avoided, since any knowledge of the 'actual' is—because of the quantum-theoretical laws—by its very nature an incomplete knowledge" (Heisenberg 1962, p. 145). I would add that this knowledge, even as concerns the "actual" (which, again, refers here to what is observed in measuring instruments), is only incomplete insofar as it cannot conform to the ideal of completeness found in classical physics, where it is accompanied by classical causality. This "incompleteness" arises because of the uncertainty relations (thus correlative to the statistical nature of quantum predictions), which apply to the actual, to the variables registered in measuring instruments and, as will be seen below, strictly speaking only to these variables. Rigorously speaking, this knowledge is as complete as nature allows our knowledge of quantum phenomena to be, as things stand now. Of course, in RWR-type interpretations, our knowledge concerning quantum objects themselves and their independent behavior is not merely incomplete but is in fact impossible. There is no knowledge or even conception concerning them at all apart from the effects they have on measuring instruments, which effects form the "actual" in Heisenberg's terminology. In other words, while there is nothing we can know or even conceive about quantum objects and behavior, we do have knowledge of and can predict, probabilistically or statistically, their effects on those aspects of the world that we can perceive, know, and think about.

The question of causality is, however, a subtle matter, especially given that one can define concepts of causality that are not classical. The subject, accordingly, merits a further discussion, which I would like to undertake now. I shall first comment, briefly, on the concepts of indeterminacy, randomness, chance, and probability, also in order to avoid misunderstanding concerning how these concepts are defined here. For they can be defined otherwise, and some among such alternative definitons could, if used instead of those given here, make some of my claims unclear or problematic.

Thus, while indeterminacy, randomness, and chance commonly refer to a manifestation of the unpredictable, their specific definitions may fluctuate. In the present definition, indeterminacy or chance is a more general category, while randomness will refer to a most radical form of indeterminacy, when even a probability is not and cannot be assigned to a possible future event. Sometimes, one speaks in such cases, of "absolute randomness." Indeterminacy (including randomness) and chance are not the same, or at least are sometimes understood differently. These differences are, however, not germane in the present context, and I shall for the sake of convenience refer to indeterminacy throughout. It may or may not be possible to estimate whether an indeterminate event would occur, or sometimes even to anticipate it as an event, in which case this event is strictly random in the present definition. An indeterminate, including random, event may or may not result from some underlying causal processes, whether this process is accessible to us or not. The first eventuality defines classical indeterminacy or randomness, conceived as ultimately underlain by a hidden classically causal architecture (which may be temporal); the second defines the irreducible indeterminacy and randomness, in the absence of such an underlying causal architecture, as is the case, automatically, in structural nonrealism. As indicated earlier, the ontological validity of the second concept of indeterminacy, including randomness, cannot be guaranteed, because one cannot definitively ascertain that this indeterminacy is not underlain by a classically causal architecture. While one might assume, as the RWR-type interpretations do, that quantum events do not result from such a process, it is impossible to definitively ascertain that they do not. It is impossible to ascertain that any indeterminate or random sequence is in fact indeterminate or random, and there is no mathematical proof that any such sequence actually is (Aronson 2013). Indeterminacy or randomness is, thus, an assumption that may only be practically justified insofar as an effective theory or interpretation based on this assumption is developed.

Although often merely glossed over, the difference between probability and statistics is important in quantum theory (QM, QFT, and QTFD). I would like to briefly comment on this difference and on the role of probability and statistics in quantum theory more generally from the present, RWR-type, perspective, which in effect makes quantum theory a form of probability theory of quantum phenomena, admittedly, an argument that has also been made while adopting a realist interpretation of quantum theory. My remarks cannot of course do justice to this vast subject and are only aimed to address a few points especially relevant for my argument here.[17] "Probabilistic" commonly refers to our estimates of the probabilities of either individual or collective events, such as that of a coin toss or of finding a quantum object in a given region of space. "Statistical" refers to our estimates concerning the outcomes of identical or similar experiments, such as that of multiple coin-tosses or repeated identically prepared experiments with quantum objects, or to the average behavior of certain objects or systems. Thus, the standard use of the term "quantum statistics" refers to the behavior of large multiplicities of identical quantum objects, such as electrons and photons, which behave differently, in accordance with, respectively, the Fermi-Dirac and the Bose-Einstein statistics. The Bayesian understanding defines probability as a degree of belief concerning a possible occurrence of an individual event on the basis of the relevant information we possess. This makes the probabilistic estimates involved, generally, subjective, at least at the time one makes them, although there may be agreement (possibly among a large number of individuals) concerning such estimates, which agreement or other exterior social factors may impact the "subjective" nature of a given estimate. The frequentist understanding, also referred to as "frequentist *statistics*," defines probability in terms of sample data by emphasis on the frequency or proportion of these data, which is considered more objective.

---

[17] For helpful comprehensive treatments, see (Khrennikov 2009) and Háyek 2014) and references there. On the Bayesian philosophy of probability, in two different versions of it, see (Jaynes 2003) and (De Finetti 2008).

In quantum physics, as noted, exact predictions are in general impossible even in dealing with elemental individual processes and events, in view of the circumstance that, while the identical preparation of the instruments used can be controlled (because the behavior of their observable parts can be described classically), the behavior of quantum objects or of quantum parts of the instruments, which enable their interactions with quantum objects, and hence these interactions, cannot be so controlled. This compelled Bohr to speak "the finite [quantum] and uncontrollable interaction" between quantum objects and measuring instruments (Bohr 1935, pp. 697, 700). This is why the outcomes of repeated identically prepared quantum experiments will in general be different. This situation could, however, be interpreted either on Bayesian lines, that is, under the assumption that a probability could be assigned to individual quantum events, or on frequentist lines, that is, under the assumption that each individual effect is strictly random. The present article, again, adopts the frequentist nonrealist, RWR-type, view, considered, in part following W. Pauli, in (Plotnitsky 2016, pp. 173-186, Plotnitsky and Khrennikov 2015).[18]

Philosophically, probability introduces an element of order into situations defined by the role of randomness in them and enables us to handle such situations better. In other words, probability or statistics is about the interplay of indeterminacy or randomness and order. This interplay takes on a unique significance in quantum physics, because of the existence of quantum correlations, such as the EPR or EPR-Bell correlations, found in the experiments of Einstein-Podolsky-Rosen (EPR) type (Einstein et al 1935) and considered in Bell's and the Kochen-Specker theorems, and numerous related findings. These correlations manifest a form of statistical order and, thus, in the present terms, structure, a structure without law. They are properly predicted by QM, a (mathematical) structure governed by a law, which is, thus, along with and responding to quantum phenomena themselves, as much about order as about indeterminacy or randomness, and, most crucially, about their unique combination in quantum physics.

I am now ready to return to causality. I shall first discuss in more detail classical causality, as an ontological category, part of the architecture of reality. The term causality has been most commonly used in this sense. It refers to the behavior of physical systems whose evolution is defined by the fact that the state of a given system (as idealized by a given theory or model) is determined at all moments of time by their state at a particular moment of time, indeed at any given moment of time. This concept is in accord with Kant's principle of causality, although the history of the principle (and of the concept of classical causality) is much longer, reaching all the way to (and even before) Plato. According to Plato: "Inquiry into nature is a search for the causes of each thing; why each thing comes into existence, why it goes out of existence, why it exists" (*Phaedo*, 96 a 6–10, Plato 2005). It is this view that was, in modern times, codified by Kant: "If, therefore, we experience that something happens, then we always presuppose that something else precedes it, which it *follows* in accordance with a rule" (Kant 1997, pp. 305, 308). This presupposition also defines the *concept* of classical causality (proceeding from causes to effects), and, in Kant and beyond, it is accompanied by and even arises from another presupposition, that of the possibility of forming a representation or at least a conception of the mechanism responsible for this rule, which presupposition defines realism. Quantum phenomena, in nonrealist, RWR-type, interpretations, violate this principle, because the cause of a given event could not in general be ascertained, even ideally or in principle. Only statistical correlations between events could be ascertained, correlations that defy classical causality.

---

[18] While neither Bohr nor Heisenberg appear to have definitively stated his position on this issue, both appear to have been inclined to a statistical view of the type adopted here. See (Plotnitsky 2016, pp. 180-184), for a discussion of Bohr's view. For a nonrealist Bayesian approach to QM, known as QBism, see (Fuchs et al 2014).

As indicated earlier, I use the concept of determinism as distinct from classical causality, a distinction that is useful for historical and conceptual reasons. If causality is an ontological category and as such is part of *the architecture of reality*, "determinism," defined as an epistemological category, is part of our *knowledge concerning reality* or of observable effects of reality. It denotes our ability to predict the state of a system, at least, again, as defined by an idealized model, exactly, rather than probabilistically, at any moment of time once we know its state at a given moment of time. Determinism is sometimes used in the same sense as classical causality, and in the case of classical mechanics, which deals with single objects or a small number of objects, both coincide.[19] Once a system is sufficiently large, one needs a superhuman power to predict its behavior exactly, as was famously explained by P. S. Laplace, who invented the figure of Laplace's demon, as an image of this power. While it follows automatically that noncausal behavior, *considered at the level of a given model*, cannot be handled deterministically, the reverse is not true. The emphasized qualification is necessary because it is possible to have *classically causal models* of those processes in nature that may not ultimately be classically causal processes. The fact that classically causal models of classical physics apply and are effective within the proper scope of classical physics does not mean that the ultimate character of the processes that are responsible for classical phenomena is classically causal. They may not be, for example, by virtue of their ultimately quantum nature. Conversely, as explained earlier, an effective model that is not classically causal, for example, that of QM in RWR-type interpretations, does not guarantee that quantum behavior itself is not causal. It may ultimately prove to be classically causal or amenable to a realist treatment, that is, a corresponding model could ultimately prove to be equally or more in accord with what is observed. Factually, quantum phenomena, again, only preclude determinism, because identically prepared quantum experiments in general lead to different outcomes: while individual quantum experiments are repeatable in terms of the state of measuring instruments before they are performed, they are not repeatable as concerns their outcomes. Only the statistics of multiple identically prepared experiments are repeatable. It would be difficult, if not impossible, to do science without being able to reproduce at least the statistical data. The lack of classical causality or of realism in the corresponding interpretations of quantum phenomena and QM are *interpretive inferences* from this situation and additional quantum features such as correlations and the uncertainty relations, or complementarity, which make it difficult or even impossible to assume the existence of a classically causal architecture underlying quantum phenomena.[20]

---

[19] There are still other uses of the term, for example, in referring to Schrödinger's equation, on which I shall comment below.

I shall now consider certain alternative conceptions of causality pertinent to my argument. Thus, the term "causality" is often used in accordance with the requirements of special relativity, which restricts (classical) causes to those occurring in the backward (past) light cone of the event that is seen as an effect of this cause, while no event can be a cause of any event outside the forward (future) light cone of that event. In other words, no physical causes can propagate faster that the speed of light in a vacuum, *c*, which requirement also implies temporal locality. Technically, this requirement only *restricts* classical causality, by a relativistic antecedence postulate, rather than precludes it, and relativity theory itself, special or general, is (locally) a classically causal and indeed deterministic theory. By contrast, while, as a probabilistic or statistical theory of quantum phenomena, QM lacks classical causality, at least in nonrealist interpretations, it respects local "causality" as concerns its probabilistic or statistical predictions, which are consistent with both temporal and spatial locality, and hence the relativistic antecedence. The same is true in the case of QTFD or QFT, which conforms to special relativity in the first place, although there are nonrelativistic versions of quantum field theory. Thus, the compatibility with relativistic or, more generally, locality requirements would be maintained insofar as an already performed experiment determines, probabilistically or (if repeated many times) statistically, a possible outcome of a future experiment, without assuming classical causality. Determinism is, again, precluded on experimental grounds. In sum, whatever *actually happens* is defined by spatially and temporally local factors, although the probabilistic or statistical *predictions*, while always *made locally* in physical terms, need not be local and could concern distant events, as in situations of the EPR type.

---

[20] Such interpretations, again, do not exclude the possibility of causal or realist interpretations of QM, or alternative causal or realist quantum theories, such as Bohmian mechanics (which is, nonlocal), or theories defined by deeper underlying causal dynamics, which makes QM an "emergent" theory. Among recent proposals is Khrennikov's "pre-quantum classical statistical field theory" (Khrennikov 2012; Khrennikov and Plotnitsky 2015).

Relativistic causality is, thus, a manifestation of a more general concept or principle, that of locality, and one can generalize relativistic causality, accordingly, without assuming special relativity first. This principle states that no instantaneous transmission of physical influences between spatially separated physical systems ("action at a distance") is allowed or that physical systems can only be physically influenced by their immediate environment. Nonlocality in this sense is usually (there are exceptions) seen as undesirable. Standard QM appears to avoid it. As just explained, under certain circumstances, such as those of the EPR-type experiments, QM can make *predictions* concerning the state of spatially separated systems, while allowing one to maintain that the physical circumstances of making these predictions and verifying them are *local*.[21] However, the question of the locality of QM or quantum phenomena is a matter of much debate and controversy, especially in the wake of the Bell and Kochen-Specker theorems and related findings, although this question was at stake in the Bohr-Einstein debate from its inception in the late 1920s, but especially following the EPR paper (Einstein et al 1935, Bohr 1935).[22]

Finally, I would like to propose the concept of quantum causality. I shall do so via Bohr's concept of complementarity, which, for reasons that will become apparent presently, Bohr saw as a generalization of classical causality, and which is consistent with relativistic causality. Complementarity is a concept defined by

(a) a mutual exclusivity of certain phenomena, entities, or conceptions; and yet
(b) the possibility of considering each one of them separately at any given point; and
(c) the necessity of considering all of them at different moments for a comprehensive account of the totality of phenomena that one must consider in quantum physics.

Complementarity may be seen as a reflection of the fact that, in a radical departure from classical physics or relativity, the behavior of quantum objects of the same type, say, electrons, is not governed, individually or collectively, by the same "physical law," especially a representational physical law, in all possible contexts, and specifically in complementary contexts. Speaking of "*physical* law" in this connection requires caution, because, in Bohr's interpretation, there is no physical law representing this behavior (and thus, again, no value-definitiveness found in classical physics or relativity), not even a probabilistic law if one adopts a statistical, rather than a Bayesian, view of the elemental individual quantum behavior. I mean that the behavior of quantum objects leads to mutually incompatible observable physical effects in complementary setups or contexts, an incompatibility not found in classical physics or relativity. On the other hand, the mathematical formalism of QM offers correct probabilistic *or* statistical predictions (no other predictions are, again, possible) of quantum phenomena *in all contexts*, in nonrealist interpretations, under the assumption that quantum objects and their behavior are beyond representation or even conception.[23]

---

[21] The EPR experiment originally proposed by EPR in dealing with continuous variables cannot be performed in a laboratory (because the quantum state defining this experiment is not normalizable), which has never put in question its legitimacy for the theoretical arguments concerning or based on it. Related experiments for discrete variables, most famously those by A. Aspect, based on Bohm's version of the EPR experiment for spin (Aspect et al. 1982), have been performed, as were experiments statistically approximating the EPR experiment.

[22] These debates cannot be addressed within the scope of this article, although the concept of "structure without law" and structural nonrealism may bear on the questions considered in these debates. The literature dealing with these subjects is nearly as immense as that on interpretations of QM. Among the standard treatments are (Bell 2004, Cushing and McMullin 1989, and Ellis and Amati 2000). See also (Brunner et al 2014), for current assessment of Bell's theorem. These theorems and most of these findings pertain to quantum data, and do not depend on QM.

If one adopts a nonrealist, RWR-type, interpretation of quantum phenomena and quantum theory, the nature of both experimental and theoretical physics changes. Experimentally we no longer track, as we do in classical physics or relativity, the independent behavior of the systems considered, track what happens in any event. Instead we define what *will* happen in the experiments we perform, by how we *experiment* with nature by means of our experimental technology, even though and because we can only predict what will happen probabilistically or statistically. Thus, in the double-slit experiment, the two alternative setups of the experiment, whether we, respectively, can or cannot know, even in principle, through which slit each particle, say, an electron, passes, we obtain two different outcomes of the statistical distributions of the traces on the screen (with which each particle collides). Or, in effect equivalently to the double-slit experiment, while also giving a rigorous meaning to the uncertainty relations, we can set up our apparatus so as to measure and correspondingly predict, again, probabilistically or statistically, either the position or the momentum of a given quantum object, but never both together. Either case requires a separate experiment, incompatible with the other, rather than representing an arbitrary selection of either type of measurement within the same physical situation, by tracking either one of its aspects or the other, as we do in classical mechanics. There, this is possible because we can, at least in principle, assign simultaneously both quantities within the same experimental arrangement. In quantum physics, we cannot. While all modern physics, from Galileo on, has always been fundamentally experimental, as it was also, and correlatively, mathematical, quantum physics changes what experiments do: they define what will or will not happen in terms of probabilistic or statistical predictions, rather than follow what is bound to happen in accordance with classical causality.[24]

---

[23] I note in passing that complementarity is connected to the concept of contextuality, prominent in discussions of quantum foundations in the wake of Bell's and the Kochen-Specker theorems and related findings, mentioned earlier, originating, again, in the EPR-type experiments. See, the works cited in note 22 and (Dzhafarov et al 2016). I might add that by being grounded in the idea of physical law, the view of complementarity just outlined is, philosophically, more radical than the view of quantum phenomena defined by the concept contextuality, *as it is commonly understood*, as in the works just cited, for example. I qualify because one could similarly ground the idea of contextuality. In effect, this grounding is implied by my outline of complementarity.

It is this probabilistic or statistical determination (which respects relativistic causality and locality) of what can or conversely cannot happen, as a result of our conscious decision concerning which experiment to perform at a given moment in time, that defines what I call "quantum causality," the concept introduced in (Plotnitsky 2016, pp. 203-206). More rigorously, quantum causality is defined as follows: Whatever happens as a quantum event and is registered as such (thus providing us with the initial data) defines a possible set of, in general probabilistically or statistically, predictable outcomes of future events and irrevocably rules out the possibility of our predictions concerning certain other, such as and in particular complementary, events.

Bohr's complementarity is clearly in accord with this concept. Indeed, this concept, in my view, gives a fitting meaning to Bohr's view of complementarity as a generalization of (classical) causality, to which it converts in the classical limit, where there is no complementarity and where our predictions could be considered ideally exact in dealing with individual or small classical systems (Bohr 1987, v. 2, p. 41). On the one hand, while stemming from "our freedom of handling the measuring instruments, characteristic of the very idea of experiment" in all physics, our "free choice" concerning what kind of experiment we want to perform acquires a new meaning with complementarity (Bohr 1935, p. 699). As against classical physics or relativity, implementing our decision concerning what we want to do will allow us to make only certain types of predictions and will exclude the possibility of certain other, *complementary*, types of predictions. Complementarity, then, generalizes causality in the absence of classical causality and, in the first place, realism, because it defines what reality can and cannot be brought about as a result of our decision concerning which experiment we perform. The predictions defined by complementarity will still be probabilistic or statistical, forming in Schrödinger's language, expectation-catalogs (Schrödinger 1935, p. 154), reflecting the probabilistic or statistical aspects of complementarity or the uncertainty relations.[25]

---

[24] Of course, we can sometimes define by an experiment what will happen in classical physics as well, say, by rolling a ball on a smooth surface, as Galileo did in considering inertia. In this case, however, we can then observe the resulting process without affecting it. This is not the case in quantum physics, because *any new observation* essentially interferes with the quantum object under investigation and defines a new experiment and a new course of events. Only *some observations* do in classical physics.

[25] If Schrödinger's equation may be seen as "deterministic," as it is sometimes (in my view, misleadingly), it is only in the sense that it *determinately* provides such expectation-catalogs.

On the other hand, while quantum causality was defined here via complementarity and while the connections between them are important, technically, quantum causality does not depend on complementarity. This independence suggests that complementary *effects*, which are comprehended by Bohr's concept of complementarity, are manifestations, in quantum phenomena, of some deeper specific, but if one adopt a nonrealist, RWR-type, view, not specifiable, ingredient of the ultimate constitution of reality. In other words, there is something in the ultimate nature of reality that is not only manifested in quantum phenomena but also in the fact that some of them are complementary. This statement is of course trivially correct if one adopts Bohr's interpretation of quantum phenomena and QM, an interpretation grounded in the concept of complementarity, although not only in this concept, given, for example, the role of his concept of phenomenon in it. My point, however, is that, as essential to the ultimate constitution of nature, this ingredient will have its effects in other, perhaps all, consistent interpretations of quantum phenomena and QM, as things stand now, including QM itself. Depending on an interpretation, these effects may be different from, if possibly related to, those captured by Bohr's concept of complementarity. Thus, the EPR-type effects, manifested in quantum correlations, may be already a manifestation of this ingredient, apart from their complementary character, which defines these effects in Bohr's interpretation (Bohr 1935, p. 700). One might, in principle, argue that the uncertainty relations (represented in the formalism of QM by the corresponding noncommuting variables) are a manifestation of this ingredient, which would imply complementarity. While the first claim is likely to be true, the second is a more complicated matter, because, although complementarity provides an interpretation of the uncertainty relations, it is not equivalent to them; and hence it is, in principle, possible to have the uncertainty relations without complementarity. Many technical and even some philosophical discussions of the uncertainty relations do not mention complementarity, although some in effect imply it. Neither Dirac's not von Neumann's classic books mention complementarity (von Neumann 1932; Dirac 1958). Feynman even saw the uncertainty relations as a remnant of classical physics, no longer necessary once one adopted a more modern view:

---

However, as *expectation*-catalogs, these catalogs are catalogs of predictions that are not deterministic, in the present view irreducibly, given the absence of classical causality in the emergence of quantum phenomena. The linearity of Schrödinger's equation or of the formalism of QM in general may be seen from this perspective, insofar as it would appear unlikely to *represent* anything physical leading to the nature of these catalogs, thus inviting an nonrealist, RWR-type, view, even if, admittedly, not making it imperative. It is enigmatic enough that QM can predict them. The RWR-type interpretations, beginning with that of Bohr, leave this quantum enigma or mystery unresolved, but they also change the nature of thought and knowledge, by making that which beyond thought and knowledge part of them.

> I would like to put the uncertainty principle in its historical place: When the revolutionary ideas of quantum physics were first coming out, people still tried to understand them in terms of old-fashioned ideas (such as, light goes in straight lines). But at a certain point the old-fashioned ideas would begin to fail, so a warning was developed that said, in effect, 'Your old-fashioned ideas are no damn good when . . . ' If you get rid of all the old-fashioned ideas and instead use the ideas that I'm expounding in these lectures—adding arrows for all the ways an event can happen [by working with the probability amplitudes]—there is no need for an uncertainty principle! (Feynman 1985, pp. 55, n. 3)

Feynman's view of the situation, expressed in here in "adding arrows for all the way an event can happen," is in fact close to quantum causality, as defined here, and in effect takes into account complementarity, which affects how these arrows can or cannot be added. It is worth noting that this type of argument would not apply to quantum correlations, which fact, however, corroborates Feynman's point. Quantum correlations are strictly quantum. In any event, although neither is *entirely* empirical (nothing, again, ever is), quantum correlations and the uncertainty relations are connected more closely to the corresponding experimental evidence and more directly predictable, probabilistically or statistically, by QM than is complementarity, which has a broader conceptual structure.[26]

I close this section, which thus far have been focused on nonrealism, with a discussion of the concept of structure. A structure is a collection, usually a set, of (abstract) objects and relationships among them, conforming to certain specified rules—a collection *organized* according certain specific rules, for example, commonly in mathematics and science, certain formal mathematical rules or, in the present extension of the concept of structure, certain correlational rules or patterns.[27] In the first case, we can define what may be called the *law* of a structure—a specifiable set of rules that govern the structure and allow one to generate new configurations within the structure or, in physics, if the structure represents independent objects and processes in nature, know or conceive, at least in principle, how such configurations emerge. In the second case, however, in dealing with (statistical) correlational structures, such may not be the case, and it is not in nonrealist interpretations of quantum correlations. There probabilities or statistics are fundamentally linked to, and are indeed defined by, the mathematical structures representing the ultimate constituents of the system considered, such as individual molecules that are assumed to behave in accordance with the laws of classical mechanics in classical statistical physics. By contrast, QM relates these two types of structures by means of predictions, in nonrealist interpretations under the assumption that there is no available or conceivable law defining the structure of these correlations themselves. There is no law that would allow us to know or even conceive of how these correlations actually emerge.

---

[26] This sense of the situation was in part suggested to me by R. B. Griffiths' recent article (Griffiths 2017), in which the deeper ingredient of reality in question appears to manifest itself in terms of "incompatible frameworks" (which are at most only indirectly related to complementarity, as defined by Bohr), and by Griffiths' comments on the first version of the present article. It is not my aim to address Griffiths' interpretation of QM (in terms of consistent histories), except by noting, by way of confirming my main point at the moment, that this interpretation is essentially different from that of Bohr or the one adopted here, which are nonrealist, RWR-type. That of Griffiths', by contrast, aims to represent, in a realist way, the ultimate constitution of reality, a representation that is, in Bohr's words, "*in principle* excluded" in Bohr's sent interpretation, the difference in fact stressed by Griffiths himself in his article.

[27] A comprehensive analysis of mathematical structuralism is given in (Corry 2013).

This situation, thus, compels us to reconceive the concept of structure. This concept still preserves the concept of organization, as it would be difficult or even impossible to conceive of a structure apart from organization and difficult to conceive of an organization apart from the law of this organization (e.g., Derrida 1978, pp. 273, 279). However, in parallel with the concept of reality without realism, this concept defined a structure as something that is without a law of structure, that is, without a conceivable law governing the emergence of this structure. While, then, it may be true that it is impossible to think of the concept of a structure lacking any law, in the present, RWR-view of quantum phenomena, this statement can be taken literally: nature, at the ultimate (quantum) level of its constitution creates (correlational) structures the emergence of which is unthinkable and as such outlaws any law that we can think of, beginning with the idea of law itself. I reiterate that at stake is not merely something that is as yet unthought, but something that is rigorously unthinkable, is beyond conception. There are, again, correlational laws or rules governing the *effects* involved and the mathematical laws of QM, within the law of the structure of the formalism, that predict these effects and these correlations, but there is no law of these correlational structures and thus no structure that would allow us to know or conceive how these correlational structures are possible.

J. A. Wheeler spoke of "law without law" in quantum theory (Wheeler 1983). It is not surprising that he ultimately linked this "law without law" to quantum information and, with the help of R. Feynman (his doctoral student), to quantum computing. Quantum processes, in their interactions with measuring instruments, create specifically organized collections of information (composed of classical bits) and make possible certain calculations, by using law-governed mathematical structures or models, but we cannot know and possibly cannot conceive how quantum processes do this. The ultimate (quantum) constitution of matter is, according to Wheeler, "it from bit," "it" inferred from "bit" (Wheeler 1990). In the present view, this "it," while real, is beyond thought, and as such, cannot ultimately be called "it," any more than anything else. Wheeler's visionary manifesto was inspired by Bohr, whom Wheeler invoked, on the same page, when he announced his "it from bit:" "The overarching principle of 20th-century physics, the quantum— and the principle of complementarity that is the central idea of the quantum — leaves us no escape, Niels Bohr tells us, from 'a radical revision of our attitude [towards the problem of] of physical reality'" (Wheeler 1990, p. 309; Bohr 1935, p. 697). (I correct Wheeler's slight misquotation of Bohr.) As I argue here, this revision is deeply rooted in Heisenberg's thinking that led him to his discovery of quantum mechanics and Bohr to his interpretation of it, which, jointly, defined the spirit of Copenhagen, as, I also argue, the spirit of structural realism and the spirit of quantum information theory. I shall now explain in more detail why this is the case.

## 4. Revisiting the Origins of Quantum Nonrealism: From Continuous Motion of Objects to Discrete States and Phenomena, with Bohr and Heisenberg

For nearly a century since the publication of von Neumann's seminal *The Mathematical Foundation of Quantum Mechanics* (Von Neumann 1932), the mathematical models of quantum theory—QM, QTFD, and QFT—have been commonly defined in terms of the Hilbert-space formalism, the axioms of which provide, in present terms, the law of this structure.[28] Grounding the quantum-mechanical formalism, already developed, if less rigorously, by Heisenberg, Born, Jordan, Dirac, and others, in this way was the aim of von Neumann's approach, with his most comprehensive exposition given in his book. By contrast, Heisenberg, in his discovery of QM and in his exposition given in his paper announcing this discovery (Heisenberg 1925), did not start with a mathematical model. He *discovered* this model, or at least a mathematical scheme, more properly developed into such a model, matrix mechanics by Born and Jordan soon thereafter (Born and Jordan 1925).[29] Heisenberg's scheme was deduced, with the help of profound, if not always fully rigorous, intuitions, from certain physical features of quantum phenomena and principles, physical and mathematical, arising from these features.[30] Other figures just mentioned primarily more properly grounded, generalized, and refined Heisenberg's scheme. Dirac offered, arguably, the most general formal version of the formalism before von Neumann, who in fact thought that Dirac's version lacked a proper mathematical rigor, in part because of Dirac's use of his delta function, which was not considered legitimate mathematically at the time. Eventually, in the 1940s, it was given a proper mathematical definition, but as a so-called "distribution," essentially a functional, by Laurent Schwartz, which retroactively mathematically legitimatized Dirac's version of quantum formalism.[31]

---

[28] While there are other versions of the quantum-mechanical formalism, such as those of C*-algebras and, more recently, category theory, thus far all equivalent mathematically, the Hilbert-space formalism remains dominant. For a helpful account on this line of thinking and its subsequent history, see (Cassinelli and Lahti 2016).

[29] It is sometimes argued that it is difficult or even impossible to properly speak of a mathematical model short of von Neumann's or some similarly axiomatic formulation. While it is, to some degree, a matter of one's definition of a model (see Note 8), this view does not appear to me justified. At least as developed by Born and Jordan, matrix mechanics, or Dirac's *q*-number formalism, is sufficient to qualify as a model. The question of what constitutes a rigorous derivation of a model is of course separate, and I will comment on it in the next section. See Notes 44 and 45 below.

[30] Schrödinger's approach to quantum mechanics was based on different, classical-like principles, but, as noted earlier, he could not avoid bringing in some quantum principles either, against his own grain (Plotnitsky 2016, pp. 84-98).

Heisenberg abandoned the project of representing the behavior of electrons in atoms. It is true that he only thought that such a representation was difficult or unlikely to achieve at the time, as opposed to arguing, as Bohr did in the 1930s, that such a representation and even an analysis, if not conception, of this behavior or quantum behavior in general was "*in principle* excluded" (Bohr 1987, v. 2, p. 62). Still, Heisenberg's was an audacious and radical move, which also decisively shaped Bohr's subsequent thinking, although Heisenberg's thinking was in turn influenced by Bohr's 1913 atomic theory. Bohr's theory only partially abandoned the geometrical representation of quantum behavior in the case of "quantum jumps"—transitions between the stationary states of electrons, conceived in terms of electrons orbiting the nuclei. While Bohr's theory, as developed by him and other had many major successes, by the early 1920s, this semi-classical view itself proved to be unsustainable. This failure compelled Heisenberg to renounce a geometrical representation of any quantum behavior, including that of stationary states in terms of orbits, and, while retaining certain key aspects of Bohr's theory, to ground his approach to quantum phenomena even more radically epistemologically. This more radical approach led him to his discovery of quantum mechanics in 1925. In this respect, to return to Einstein's characterization, "Heisenberg's purely algebraic method" was a shift from geometry to algebra in fundamental physics, a shift, in this case, correlative to those toward the concepts of structure without law, reality without realism, and probability without causality, none of which would have appealed to Einstein, because there was no longer an algebraic representation either, continuous or discontinuous, of quantum objects and behavior either (Einstein 1936, p. 378).

It is true that we speak of, and can rigorously define, Hilbert *spaces* and their *geometry*, and in this sense, one could also speak, as Dirac was reportedly fond of doing, of geometrical thinking in QM, by analogy with projective or finite geometries, which equally defy visualization by means of our general phenomenal intuition. This, however, is not in conflict with my point, or with Einstein's characterization of Heisenberg's method as essentially algebraic. I would argue that these spaces and geometries may, as spaces and geometries, be seen as "symbolic" or "metaphorical," or better, in terms of structure, essentially, an algebraic concept. Ultimately, these spaces and their geometries are special,

---

[31] This case poses deep questions concerning the mathematical rigor of physics or of mathematical models used in physics, and it became famous in this connection. See, Note 29 above.

"spatial-like," forms of *algebra*, as against the more standard forms of geometry, such as Euclidean or even non-Euclidean geometry, or differential geometry, used in general relativity. The geometries of these more conventional spaces are, however, defined by certain algebraic properties and relations (such as "distance"), some of which can be used to define Hilbert spaces. In other words, these properties and relations define the *structure* of Hilbert spaces or, similarly, other mathematical "spaces," (such as those of projective geometries, abstract algebraic varieties, the spaces of noncommutative geometry, geometric groups, and so forth), but in the absence of certain other, more conventionally spatial elements and structures, geometrical or topological, found in more conventional spatial objects, such as the three-dimensional Euclidean space, the paradigmatic model of phenomenal spatiality. This is why I speak of the symbolic or metaphorical character of "space" and "geometry" in the case of Hilbert spaces. This character arises in part by virtue of the infinite-dimensional nature of some of those "spaces" or also the fact that they are defined over complex numbers, which appear irreducible in QM. The Hilbert spaces involved are finite in the case of discrete variables, but these spaces can still have higher (if finite) dimensions and are over complex numbers, and thus are ultimately "algebraic" in the same sense. It is difficult to surmise, especially from reported statements, what Dirac exactly had in mind in his appeals to geometrical thinking in quantum theory. If, however, one is to judge by his writings, they appear to suggest that at stake are the algebraic properties and relations modeled on those found in geometrical objects, defined by algebraic structures, as just explained, such as, at the time, projective geometry. Indicatively, notwithstanding his insistence on geometry in Dirac's quantum-theoretical thinking, O. Darigold's analysis of this thinking shows precisely the significance of algebra there (Darigold 1993). Thus, as he says, "roughly, Dirac's quantum mechanics could be said to be to ordinary mechanics what noncommutative geometry is to intuitive geometry" (Darigold 1993). Noncommutative geometry, the invention of which was in part inspired by the mathematics of QM, is in effect a form of spatial-like algebra just described (Connes 1994, p. 38; Plotnitsky 2009, pp. 112-113).

In what sense, apart from certain essentially algebraic properties, may such spaces be seen as spaces, for example and in particular, in the sense of visualizing, imagining, or intuiting them? The subject is complex, and it is, it might be added, still far from sufficiently explored in cognitive psychology and related fields. It is almost certainly true that, when we visualize such objects, we visualize only three- (and even two-) dimensional configurations and supplement them by algebraic structures and intuitions, as Feynman observed in describing visual intuition in thinking about quantum objects (cited in Schweber 1994, pp. 465–466). These problems, again, manifested in finite and projective geometries, arise even for spaces or geometries of finite dimensions, once the number of dimensions is more than three, or for spaces of any dimensions, beginning with the complex plane itself, defined over number fields other than that of real numbers, such as those used in complex analysis or algebraic geometry, or, again, noncommutative geometry. Algebraic geometry even arrives at such esoteric concepts as spaces without points. Discrete geometries (which have the topological dimension zero), introduced in the nineteenth century, or projective geometries, and even non-Euclidean geometries, pose these problems as well. J. Tate in this connection famously defined the following philosophy, which fully applies in spaces and geometries of quantum theory: "Think Geometrically, Prove Algebraically:"

> It is also possible to look at polynomial equations and their solutions in rings and fields other than **Z** or **Q** or **R** or **C**. For example, one might look at polynomial with coefficients in the finite field $F_p$ with $p$ elements and ask for solutions whose coordinates are also in the field $F_p$. You may worry about your geometric intuitions in situations like this. How can one visualize points and curves and directions in $\mathbf{A}^2$ when the points of $\mathbf{A}^2$ are pairs $(x, y)$ with $x, y \in F_p$? There are two answers to this question. The first and most reassuring is that you can continue to think of the usual Euclidean plane, i.e., $\mathbf{R}^2$, and most of your geometric intuitions concerning points and curves will still be true when you switch to coordinates in $F_p$. The second and more practical answer is that the affine and projective planes and affine and projective curves are defined *algebraically* in terms of ordered pairs $(r, s)$ or homogeneous triples $[a, b, c]$ without any reference to geometry. So in proving things one can work algebraically using coordinates, without worrying at all about geometrical intuitions. We might summarize this general philosophy as: *Think Geometrically, Prove Algebraically*" (Silverman and Tate 2015, p. 277).

Rigorously then, in these cases, or in dealing with Hilbert spaces in quantum mechanics, and in truth in the cases of most so-called spatial and geometrical *structures* we ultimately think algebraically, too, even if with the help of geometrical or topological intuitions. It is quite obvious that these intuitions are ultimately limited even when we deal with curves in the usual Euclidean plane, let alone in considering objects like Riemann surfaces as curves over **C**.

In sum, Einstein was quite right in characterizing Heisenberg's method as "purely algebraic." The "geometry" of the Hilbert spaces of quantum theory only confirms the algebraic character of this method, admittedly especially manifested in his matrix or operator algebra. These algebraic structures, as structures defined by law, relate to structures without law manifested in quantum phenomena, in the absence of any underlying ultimate geometrical reality, akin to that found in relativity, for example, which was of course what was most disconcerting to Einstein in "Heisenberg's purely algebraic method."

Although one could sometimes surmise the meaning of the term "principle" from its use, it is, similarly to the term "concept," rarely defined or explained in physical or even philosophical literature. His title notwithstanding, Heisenberg did not do so in his first book, *The Physical Principles of the Quantum Theory* (Heisenberg 1930), nor did Dirac, in his equally famous, *Principles of Quantum Mechanics*, originally published in the same year as Heisenberg's book (Dirac 1930). For present purposes, I shall adopt the concept of principle from Einstein's concept of a "principle theory," which he introduced by way of juxtaposing this concept to that of a "constructive theory." This concept corresponds to the use of principles by Bohr and Heisenberg, and by quantum-information theorists discussed in the next section. According to Einstein, constructive theories aim "to build up a picture of the more complex phenomena out of the materials of a relatively simple formal scheme from which they start out," which, it follows, also make such theories realist. By contrast, principle theories "employ the analytic, not the synthetic, method. The elements which form their basis and starting point are not hypothetically constructed but empirically discovered ones, general characteristics of natural processes, principles that give rise to mathematically formulated criteria which the separate processes or the theoretical representations of them have to satisfy" (Einstein 1919, p. 228). I would add the following qualification, which is likely to have been accepted by Einstein: Principles are not empirically discovered but formulated on the basis of empirically established evidence. A principle theory may also be a constructive theory, but it need not be, and Heisenberg's matrix mechanics was not.[32]

---

[32] It should be noted that, although the terms "principle," "postulate," and "axioms," are sometimes used in physics indiscriminately and interchangeably with each other (mathematicians, by necessity, tend to be more careful), the term and concept of "principle" is used in this article strictly in the specific definition just given. It is difficult to entirely avoid overlapping between the concepts designated by these terms, and sometimes those designated as "laws," especially because physical principles are often accompanied by or derive from (or conversely give rise to) postulates. It may also be a matter of the functioning of these concepts. Thus, conservation laws are sometimes seen as conservation principles. It is possible, however, to sufficiently, even if not entirely, analytically separate these concepts, along the following lines. Euclid and, it appears, the ancient Greeks in general, distinguished between "axioms" and "postulates." Axioms were thought to be something manifestly self-evident, such as the first axiom of Euclid ("things equal to the same thing are also equal to each other"). A postulate, by contrast, is *postulated*, in the sense of "let us assume that *A* holds," thus indicating primarily that one aims to proceed under assumption *A* and see what follows from it according to established logical rules (this is similar to proceeding from axioms), rather than claiming *A* to be a self-evident truth. Euclid's postulates may be thought of as those assumptions that were necessary and sufficient to derive the truths of geometry, of some of which we might already be intuitively persuaded (e.g., "the first postulate: to draw a straight line from any point to any point"). The famous fifth postulate is a case in point. It defines Euclidean geometry alone, which in part explains millennia of attempts to derive it as a theorem. Keeping in mind further complexities potentially involved in using and defining these concepts in geometry and beyond, I shall adopt this Euclidean understanding of axioms and postulates. Given that my subject is physics, I shall primarily refer to postulates, assumed on the basis of experimental evidence (as it stands now and hence is potentially refutable) and often, but not always, grounding principles. This is, to some degree, in contrast to the axiomatic approaches to quantum foundations, which, from von Neumann

Heisenberg's approach and then Bohr's interpretation of QM adopted the following principles (with Bohr's complementarity principle added in 1927):

(1) the principle of discreteness or the QD principle, according to which all observable quantum phenomena are individual and discrete in relation to each other (which is not the same as the atomic discreteness of quantum objects themselves);

(2) the principle of the probabilistic or statistical nature of quantum predictions, the QP/QS principle, which is maintained (in contrast to classical statistical physics) even in the case of elemental individual quantum processes, and is, as a principle, accompanied by a special, nonadditive, nature of quantum probabilities and rules, such as Born's rule, for deriving them; and

(3) the correspondence principle, which, as initially understood by Bohr, required that the predictions of quantum theory must coincide with those of classical mechanics at the classical limit, but which was given by Heisenberg a more rigorous form and was made into "the mathematical correspondence principle," requiring that the equations of QM convert into those of classical mechanics in the classical limit.

The QD principle originated in the quantum postulates of Bohr's 1913 atomic theory, according to which, "the frequencies [of emitted radiation] appear as functions of two variables … in the form:

$$v(n, n - \alpha) = 1/h \{W(n) - W(n - \alpha)\}$$" (Heisenberg 1925, p. 263).

This rule and Bohr's quantum postulates, to begin with, are in irreconcilable conflict with both classical mechanics and classical electrodynamics. The QD and QP/QS principles are in fact correlative. The QP/QS principle is expressed in the quantum-mechanical formalism by the formula for the probability amplitudes cum Born's rule, which was a postulate added to, rather than derived from, the formalism. Heisenberg only formulated or (given that it was not expressly stated by him) used this postulate in the special case of the transitions

---

on, are primarily, even if not entirely (it is not possible to avoid physical postulates or principles), concerned with the mathematical formalism of QM, where the term axiom is more fitting. Next to nothing has the self-evidence of axioms even in classical mechanics (in part, as against, its mathematical models), and most of the uses of the term "axiom" in physics, including in the quantum-informational schemes considered here, are in effect closer to that of "postulate," as just defined. Bohr, more careful and more etymologically attuned than most in using his terms, prefers both postulates, such as the quantum postulate (Bohr 1987, v. 1, pp. 52-53), and principles, such as his correspondence principle. Complementarity functions as both a concept and a principle in Bohr. The term was also eventually used by Bohr to designate his overall interpretation of quantum phenomena and quantum mechanics.

between stationary states ("quantum jumps"), and not, as Born did, as universally applicable in QM. Referring to stationary states requires cautions, because *stationary*, in Bohr's theory, only means that the electrons remained in their orbits with the same energy, were in the same "energy-state," but would continuously change their position or their "position-state" along each orbit. On the other hand, the electrons would discontinuously, by "quantum jumps," change their energy states, or their other states, by moving from one orbit to another. In Heisenberg, there were no longer orbits but only states and discontinuous transitions between states.

As noted from the outset, one was no longer thinking, as in classical mechanics, in terms of predictions, even probabilistic predictions, concerning *a moving object*, say, an electron, free or orbiting the nucleus of an atom, but instead in terms of the probabilities of *transitions between the states of an electron* (Freidel 2016). This would imply that the concept of motion, as we know it (classical or relativistic), could no longer apply to an electron, but some form of concept of physical state, such as that corresponding to a given energy measurement, could be, albeit with qualification and caution, if one assumes that the independent behavior of the atom is beyond representation or even conception, including a mathematical one.[33] I shall comment on this point presently. This type of thinking itself emerged already in Bohr's 1913 atomic theory in considering an electron's transitions from one energy level to another (levels corresponding to stationary states) but became a new way of thinking about phenomena or objects in quantum physics in general. As Heisenberg said shortly before he completed his paper introducing QM: "What I really like in this scheme is that one can really reduce *all interactions* between atoms and the external world ... *to transition probabilities*" [between states] (W. Heisenberg, Letter to Kronig, 5 June 1925; cited in Mehra and Rechenberg 2001, v. 2, p. 242; emphasis added).[34]

Heisenberg's scheme, thus, extended Bohr's 1913 concept of discrete transitions, "quantum jumps," which had no mechanical or geometrical model, between stationary states, which had a mechanical and geometrical model and were conceived by Bohr as elliptical orbits on the planetary model in classical physics. "This concept," as Heisenberg noted later, "had been somewhat doubtful *from the beginning*," because of "the discrepancy between the calculated orbital frequency of the electrons and the frequency of the emitted radiation," which had to be interpreted as a limitation to the concept of the electronic orbit" (Heisenberg 1962, p. 41; emphasis added). This observation, made in the late 1950s, may represent more Heisenberg's own thinking around the time of his creation of quantum mechanics in 1925, than Bohr's earlier thinking. If anything, it was Bohr's idea of quantum jump that appeared doubtful to most when Bohr introduced his theory. Be that as it may on this score, accepting this discrepancy and, thus, dissociating these two types of frequencies was a revolutionary move on Bohr's part, emphasized as such by Bohr himself. As he said in his first paper presenting his theory, "How much the above interpretation differs from an interpretation based on the ordinary electrodynamics is perhaps most clearly shown by the fact that we have been forced to assume that a system of electrons will absorb radiation of a frequency different from the frequency of vibration of electrons calculated in the ordinary way" (Bohr 1913, p. 149).

---

[33] I am not referring here to the concept of quantum state as a state vector in a Hilbert space, which is, in the present view, merely part of the predictive machinery of quantum mechanics and has no physical meaning in itself.

[34] More classical views of the situation have persisted as well, at the earlier stages of QM especially following Schrödinger's introduction of his wave mechanics, which appeared more amenable to such views. The debates concerning the subject have never subsided and were rekindled by Bell's and the Kochen-Specker theorems, and related findings.

In retrospect, Bohr's 1913 postulates almost cried out: "Give up the idea of orbits!" Heisenberg did just that. He rethought stationary states as just energy states, permitting no mechanical model or geometrical representation either, thus excluding from his theory the mechanical picture of both stationary states and transitions between them, manifested in measurable quantities and changes in the values of these quantities. In sum, he excluded both altogether. He was also able, as his greatest achievement, to give a workable mathematics to this physics, the mathematics of quantum mechanics. This took another decade, however. A retrospective view, while not without its benefits, is rarely a reliable guide to how discoveries occur. A few things had to happen before the cry "give up the idea of orbits" could have been heard or, for that matter, even made, because the dominant sentiment at the time was that a return to a more classical picture of the atomic constitution was likely. While the hope for this return, personified by Einstein, has remained alive among physicists and philosophers, and is still as fervent as ever even now, before Heisenberg the type of theory he proposed was all but inconceivable. However, although Bohr's theory had, in the hands of Bohr and others, many impressive successes to its credit during the decade following its introduction, the idea of orbits for stationary states ran into major and ultimately insurmountable difficulties. On the other hand, the approach to quantum theory as a (probabilistic or statistical) theory predicting the transitions between states was working well in Bohr's theory itself and beyond, as in Einstein's remarkable 1916 treatment, using Bohr's theory, of spontaneous and induced emission and absorption of radiation (Einstein 1916a, b; Freidel 2016). So, the situation became ripe for giving up the concept of an orbit in considering the behavior of the electrons in atoms, at least ripe for Heisenberg. Not everyone was ready to do so, even among those whom one would have expected to. Thus, as is clear from Heisenberg's letter to him, Pauli, who did question the idea of orbits previously, appears to have failed to completely renounce it at the time of Heisenberg's discovery, although he did so soon thereafter:

> But I do not know what you mean by orbits that fall into the nucleus. We certainly agree that already the kinematics of quantum theory is totally different from that of classical theory ($h\nu$-relations), hence I do not see any geometrically-controllable sense in the

> statement "falling into the nucleus." It is really my conviction that an interpretation of the Rydberg formula in terms of circular and elliptical orbits (according to classical geometry) does not have the slightest physical significance. And all my wretched efforts are devoted to killing totally the concept of an orbit—which one cannot observe anyway—and replace it by a more suitable one. (Heisenberg to Pauli, 9 July 1925; cited in Mehra and Reihenberg 2001, v. 2, p. 284; emphasis added)

The reference to "classical geometry" is worth noting, in juxtaposition to Heisenberg's "algebraic method," bemoaned by Einstein. The concept of orbits remained applicable and useful for the higher orbits, when "electrons should move at a large distance from the nucleus just as they do when one sees them moving through a cloud chamber," in effect, a manifestation of the correspondence principle, very effectively used, in its mathematical form, by Heisenberg in his scheme (Heisenberg 1962, p. 41). It should be kept in mind that the electrons' behavior is still quantum in these regions and, hence, the concept of motion is a classical approximation, ultimately possible because one could only observe the effects of this behavior in the measuring instruments. On the other hand, certain quantum effects, which are not registered when we deal with classical objects, could still in principle be observed, albeit with a low probability.

Indeed, Heisenberg's statement in his letter to Kronig cited above also suggests that his new QM was about the *interactions* between atoms in the observed external world, specifically the measuring instruments involved (or those objects in nature that could be used as instruments), a view manifested in Heisenberg's paper. All that one could say about quantum objects and processes could only concern their effects on measuring instruments, effects probabilistically or statistically predictable by means of QM. In this sense, as explained earlier, one can speak of a physical state of a quantum object, such as an electron, only insofar as a certain change is registered in the measuring instruments associated with this electron, say a registered change in two energy levels, which we associate with the corresponding stationary states, or a registered change in the position state of a free electron. As noted earlier, while it is reasonable to assume that something has "changed" or "happened" between these two registered events, one must, especially if one adopts the RWR-type view, keep in mind, that, to return to Heisenberg's formulation, "these concepts cannot be applied in the space between the observations; they can only be applied at the points of observation" (Heisenberg 1962, p. 145).[35] We certainly have no means to

---

[35] J. Barbour's concept of "Platonia," an underlying reality without change and motion (the idea originating with Parmenides) appears to derive from this circumstance (Barbour 1999). From the present, RWR-principle-based, viewpoint, however, it does not follow that everything "stands still," at the ultimate level of reality, because the concept of reality without change and motion would not apply any more than that of change or motion. It only follows that no human concept of time or

ascertain where exactly and when exactly this change occurred: it always happens, in Lucretius's phrase (used by him, admittedly, in the classical picture of the swerve of an atom from a given trajectory), "at quite uncertain times, and uncertain places" (Lucretius 2009, Book Two, ll. 218–219, p. 42).[36] Nor, again, do we have means to repeat any such experiment exactly, but could only have the verifiable statistics of many repeated experiments, each of which will generally lead to a different outcome.

The mathematical correspondence principle motivated Heisenberg's decision to retain the equations of classical mechanics, while, necessarily, introducing mathematically different variables. Because these variables were different, the correspondence with classical theory was defined by the fact that new quantum variables could be substituted for by conventional classical variables (such as those of position and momentum) in the classical limit, when, as explained above, the electrons were far away from the nuclei and when, accordingly, a classical concept, such as orbits, could be retained, even though the electrons' behavior was still quantum. While using the correspondence principle (in its original ad hoc way), the old quantum theory was defined by the strategy of retaining the variables of classical mechanics while adjusting the equations to achieve better predictions. Heisenberg's reversal of this strategy was, thus, unexpected as well. This reversal, however, required a radical change of the role these equations were to play. They no longer represented the motion of the objects considered, but instead served as mathematical means enabling probabilistic or statistical predictions concerning effects (in this case spectra, assumed to result from the emission of radiation by electrons) manifested in measuring instruments.

Heisenberg's discovery was a remarkable achievement, ranked among the greatest in the history of physics.[37] A detailed discussion of his derivation of QM is beyond my scope.[38] Several key features of his thinking are, however, worth commenting on. Heisenberg's new quantum variables, to which the equations of QM (again, formally the same as those of classical mechanics) applied, were infinite unbounded matrices with complex elements. Their multiplication, which Heisenberg, who was famously unaware of the existence of matrix algebra and reinvented it, had to define, is in general not

---

space (and there is no other such concept than human) would be applicable either. Heisenberg certainly would not subscribe to Barbour's argument.

[36] See (Rovelli 1998) for a different viewpoint on this situation.

[37] This does not mean that Heisenberg's invention of QM was independent of, or was not helped by, preceding contributions, even beyond the key pertinent works in the old quantum theory by Einstein, Bohr, Sommerfeld, and others. H. Kramers's work on dispersion and his collaboration with Heisenberg on the subject were especially important for Heisenberg's work. See (Mehra and Rechenberg 2001, v. 2) for this history.

[38] I have considered this derivation in detail on previous occasions, most recently in (Plotnitsky 2016, pp. 68-83).

commutative. It is not even entirely clear whether Heisenberg even thought of these variables as matrices, as opposed to, in his words, just "ensembles of quantities" (Heisenberg 1925, p. 264). Their properly matrix nature was realized by M. Born, who, together with P. Jordan, developed Heisenberg's scheme into matrix mechanics (Born and Jordan 1925). Essentially, these variables are operators in Hilbert spaces over complex numbers, although Heisenberg did not know this at the time either. Such mathematical objects had never been used in physics previously, and their noncommutative nature was, initially, questionable and even off-putting for some, including Heisenberg himself and Pauli, both of whom, however, quickly changed their views on this point (Plotnitsky 2009, pp. 90, 111, 116).[39] In fact, while matrix algebra, in both finite and infinite dimensions, was developed in mathematics by then, *unbounded* infinite matrices were not previously studied. As became apparent later, such matrices are necessary to derive the uncertainty relations for continuous variables.

Heisenberg's invention of his matrices was made possible by the idea of arranging algebraic elements corresponding to numerical quantities (transition probabilities or, more accurately, probability density functions) into infinite square tables, at least in effect, given the double-indexing of these elements. It is true that, once one deals with *transitions* between stationary states and their energy levels, matrices appear naturally, with rows and columns defined by these transitions. This naturalness, however, became apparent or, one might say, *became natural*, only in retrospect. This arrangement was the phenomenological construction of a mathematical object, a matrix, considered as an element of a mathematical structure, part of (infinite-dimensional) linear algebra, essentially, again, an operator algebra, now known as that of "observables," in a Hilbert space over complex numbers. There are further details: for example, as unbounded self-adjoint operators, defined on infinite dimensional Hilbert spaces, these matrices do not form an algebra with respect to the composition as a noncommutative product, although some of them satisfy the canonical commutation relation. These details are, however, secondary here. Most crucial was that the concept was used physically in a fundamentally different way from the way in which representational concepts of classical physics or relativity were used. Heisenberg's variables were mathematical entities enabling probabilistic or statistical predictions concerning the relationships between *quantum phenomena*, observed in measuring instruments, without providing a mathematically idealized representation of the behavior of the *quantum objects* responsible for the appearance of these phenomena.

---

[39] Technically, tensors of the second rank, used in relativity, are matrices and thus are, in general, noncommutative. However, their role is entirely different that of from matrices or operators in QM, where they replace, as complex-valued variables in the equations of the quantum-mechanical formalism, real-valued functions used in the equations of classical physics to represent (in realist mathematical models) physical variables, such as position, momentum, and so forth.

In this regard, although understandable historically, the term "observables" may be misleading and is especially inadequate if one adopts a nonrealist view. Beginning with his response to Heisenberg's discovery of QM, Bohr saw the quantum-mechanical formalism as "symbolic" in the following sense. While the mathematical symbols used in it appear, as variables, in the same equations as those used in classical mechanics, these symbols did not represent physical quantities pertaining to quantum objects themselves and their behavior, in the way such symbols do in classical mechanics. By the same token, the equations of QM, Schrödinger's equation included, no longer function as equations of motion, at least in the sense of classical physics or relativity. Instead they are part of the probabilistic mathematical technology of QM, enabling us to compile, in Schrödinger's terms, expectation-catalogs concerning possible future quantum events (Schrödinger 1935a, p. 154).

Heisenberg began his derivation with an observation that reflects a radical departure, which he saw as necessary, from the classical ideal of continuous mathematical representation of individual physical processes in dealing with discrete quantum events, still using continuous mathematics to do so, but no longer in a representational, but only probabilistically predictive, way. He says: "in quantum theory it has not been possible to associate the electron with a point in space, *considered as a function of time*, by means of observable quantities. However, even in quantum theory it is possible to ascribe to an electron the emission of radiation" [the effect of which emission could be observed in a measuring instrument] (Heisenberg 1925, p. 263; emphasis added). My emphasis reflects the fact that, in principle, a measurement could associate an electron with a point in space, but not by linking this association to a function of time representing the continuous motion of this electron, in the way it is possible in classical mechanics.[40] If one adopts an RWR-type interpretation, one cannot assign any properties to quantum objects themselves, not even single such properties, such as that of having a position, rather than only certain joint ones, which are precluded by the uncertainty relations. One could only assign physical properties to the measuring instruments involved. Heisenberg described his next task as follows, which, again, shows the genealogy of his derivation in Bohr's atomic theory, through the QD principle: "In order to characterize this radiation we first need the frequencies which appear as functions of two variables. In quantum theory these functions are in the form:

$$v(n, n - \alpha) = 1/h \{W(n) - W(n - \alpha)\}$$

---

[40] Matrix mechanics did not offer a treatment of stationary states, in which and only in which one could in principle speak of the position of an electron in an atom, although while there are stationary (energy) states, an electron itself is never stationary, is always in motion.

and in classical theory in the form

$$v(n, \alpha) = \alpha v(n) = \alpha/h(dW/dn)$$" (Heisenberg 1925, p. 263).

This difference leads to a difference between classical and quantum theories as regards the combination relations for frequencies, which, in the quantum case, correspond to the Rydberg-Ritz combination rules, again, reflecting, to return to Heisenberg's locution, "the discrepancy between the calculated orbital frequency of the electrons and the frequency of the emitted radiation." However, "in order to complete the description of radiation [in correspondence, by the mathematical correspondence principle, with the classical Fourier representation of motion] it is necessary to have not only frequencies but also the amplitudes" (Heisenberg 1925, p. 263). On the one hand, then, by the correspondence principle, the new, quantum-mechanical equations must formally contain amplitudes, as well as frequencies. On the other hand, these amplitudes could no longer serve their classical physical function (as part of a continuous representation of motion) and are instead related to discrete transitions between stationary states. (Nor ultimately do frequencies because of the non-classical character of the Rydberg-Ritz combination rules.) In Heisenberg's theory and in QM since then, these "amplitudes" are no longer amplitudes of physical motions, which makes the name "amplitude" itself an artificial, *symbolic* term. In commenting on linear superposition in quantum mechanics in his classic book, Dirac emphasized this difference: "*the superposition that occurs in quantum mechanics is of an essentially different nature from any occurring in the classical theory*" (Dirac 1958, p. 14). In nonrealist interpretations, this superposition is not even physical: it is only mathematical. In classical physics this mathematics represents physical processes; in quantum mechanics, at least in the nonrealist view, it does not. Amplitudes are instead linked to the probabilities of transitions between stationary states: they are essentially what we now call probability amplitudes. The corresponding probabilities are derived, from Heisenberg's matrices, by a form of Born's rule for this limited case. (As I said, technically, one needs to use the probability density functions, but this does not affect the essential point in question.) The standard rule for adding the probabilities of alternative outcomes is changed to adding the corresponding amplitudes and deriving the final probability by squaring the modulus of the sum. This reconceptualization of "amplitude" is an extension of the conceptual shift from finding the probability of finding an electron in a given state to the probability of the electron's discrete transitions ("quantum jumps") from one state to another, found in Bohr's theory and manifested in Bohr's frequency rule.

One can also see here, as part of Heisenberg's "purely algebraic method," a conversion of the classical continuous geometrical picture of oscillation or wave propagation, as defined by frequencies and amplitudes, into the algebra of probabilities of transitions between discrete quantum events. This conversion is accomplished by redefining amplitudes, as just explained. Schrödinger tried to "restore" a wave-type continuous geometry to quantum theory in his wave mechanics, but he could not successfully overcome the irreducible nature of quantum discreteness, manifested in the QD principle. His formalism itself can be (and quickly was) interpreted on Heisenberg's lines under discussion, which is not surprising given the mathematical equivalence of both schemes, established shortly thereafter, by Schrödinger, among others.

The mathematical structure thus emerging is in effect that of vectors and (in general, noncommuting) Hermitian operators in complex Hilbert spaces, which are infinite-dimensional, given that one deals with continuous variables. Heisenberg explains the situation in these terms in (Heisenberg 1930, pp.111-122). In his original paper, which reflect his thinking more directly, he argues as follows:

> The amplitudes may be treated as complex vectors, each determined by six independent components, and they determine both the polarization and the phase. As the amplitudes are also functions of the two variables $n$ and $\alpha$, the corresponding part of the radiation is given by the following expressions:
> Quantum-theoretical:
>
> $$\text{Re}\{A(n, n-\alpha)e^{i\omega(n, n-\alpha)t}\}$$
>
> Classical:
>
> $$\text{Re}\{A_\alpha(n)e^{i\omega(n)\alpha t}\}. \text{ (Heisenberg 1925, p. 263)}$$

The problem—a difficult and, "at first sight," even insurmountable problem—is now apparent: "the phase contained in $A$ would seem to be devoid of physical significance in quantum theory, since in this theory frequencies are in general not commensurable with their harmonics" (Heisenberg 1925, pp. 263-264). This incommensurability, which is in an irreconcilable conflict with classical electrodynamics, was, again, one of the most radical features of Bohr's 1913 atomic theory, on which Heisenberg builds here. His strategy is still based on the shift from calculating the probability of finding a moving electron in a given state to calculating the probability of an electron's transition from one state to another, without describing the physical mechanism responsible for this transition. Heisenberg's theory is even more in harmony with this approach because there are no longer orbits, where the preceding approach would still apply. Just as Bohr before him, Heisenberg converts this, to the classical way of thinking, insurmountable problem into a possible solution, defined by and defining a new way of thinking. Heisenberg invents QM by showing that the real problem here is classical way of thinking, because it is not in accord with how nature or, more accurately, our interactions with nature work when dealing with the atomic constitution of nature. One needs a new way of theoretical thinking to restore this accord.

Heisenberg says next: "However, we shall see presently that also in quantum theory the phase has a definitive significance which is *analogous* to its significance in classical theory" (Heisenberg 1925, p. 264; emphasis added). "Analogous" could only mean here that, rather than being analogous physically, the way the phase enters mathematically is analogous to the way the classical phase enters mathematically in classical theory, in accordance with the *mathematical* form of the correspondence principle, insofar as quantum-mechanical equations are formally the same as those of classical physics. Heisenberg only considered a toy model of an aharmonic quantum oscillator, and thus needed only a Newtonian equation for it, rather than the Hamiltonian equations required for a full-fledged theory, developed by Born and Jordan (Born and Jordan 1925; Born, Heisenberg, and Jordan 1926). As Heisenberg explains, if one considers "a given quantity *x(t)* [a coordinate as a function of time] in classical theory, this can be regarded as represented by a set of quantities of the form

$$A_\alpha(n) e^{i\omega(n)\alpha t},$$

which, depending on whether the motion is periodic or not, can be combined into a sum or integral which represents *x(t)*:

$$x(n, t) = \sum_\alpha{}_{-\infty}^{+\infty} A_\alpha(n)\, e^{i\omega(n)\alpha t}$$

or

$$x(n, t) = \int_{-\infty}^{+\infty} A_\alpha(n)\, e^{i\omega(n)\alpha t}\, d\alpha\text{''} \quad \text{(Heisenberg 1925, p. 264)}.$$

Heisenberg next makes his most decisive and most extraordinary move. He notes that "a similar combination of the corresponding quantum-theoretical quantities seems to be impossible in a unique manner and therefore not meaningful, in view of the equal weight of the variables *n* and *n* − *α*" (Heisenberg 1925, p. 264). "However," he says, "one might readily regard the ensemble of quantities $A(n, n - \alpha)e^{i\omega(n, n - \alpha)t}$ [an infinite square matrix] as a representation of the quantity *x(t)*" (Heisenberg 1925, p. 264). The arrangement of the data into these ensembles, in effect square tables, is a brilliant and, as I said, in retrospect, but only in retrospect, natural way to connect the relationships (transitions) between stationary states. However, it does not by itself establish an *algebra* of these arrangements, for which one needs to find the rigorous rules for adding and multiplying these elements.

Otherwise Heisenberg cannot use these variables in the equations of his new mechanics. To produce a quantum-theoretical version of the classical equation of motion that he considered, which would apply (no longer as an equation of motion!) to these new variables, Heisenberg needs to be able to construct the powers of such quantities, beginning with $x(t)^2$, which is actually all that he needs for his equation. The answer in classical theory is obvious and, for the reasons just explained, obviously unworkable in quantum theory. Now, "in quantum theory," Heisenberg proposes, "it seems that the simplest and most natural assumption would be to replace classical [Fourier] equations … by

$$B(n, n-\beta)e^{i\omega(n, n-\beta)t} = \sum_{\alpha=-\infty}^{+\infty} A(n, n-\alpha)A(n-\alpha, n-\beta)e^{i\omega(n, n-\beta)t}$$

or

$$= \int_{-\infty}^{+\infty} A(n, n-\alpha)A(n-\alpha, n-\beta)e^{i\omega(n, n-\beta)t}\, d\alpha\text{"} \text{ (Heisenberg 1925, p. 265).}$$

This is the main mathematical postulate, the (matrix) multiplication postulate, of Heisenberg's new theory, "and in fact this type of combination is an almost necessary consequence of the frequency combination rules" (Heisenberg 1925, p. 265). This combination of the particular arrangement of the data and the construction of an algebra of multiplying his new variables is Heisenberg's great invention. (As I noted, technically, these matrices do not form an algebra with respect to the composition as a noncommutative product.) The "naturalness" of this construction or Heisenberg's claim that it is "an almost necessary consequence of the frequency combination rules" should not hide the radical and innovative nature of Heisenberg's discovery, one of the greatest in twentieth-century physics. If anything, these are the frequency combination rules that become a necessary consequence of this construction.

Although it is commutative in the case of squaring a given variable, $x^2$, this multiplication is in general noncommutative, expressly for position and momentum variables, and Heisenberg, without quite realizing, used this noncommutativity in solving his equation, as Dirac was the first to notice. Heisenberg was of course aware that his new variables, in general, did not commute, although the significance of this fact became fully apparent only with Born and Jordan's paper (Born and Jordan 2015). Taking his inspiration from Einstein's "kinematics" of special relativity, Heisenberg spoke of his new algebra of matrices as the "new kinematics." This was not the best choice of term because his new variables no longer described or were even related to motion as the term kinematic would suggest, one of many, historically understandable, but potentially confusing terms. Planck's constant, *h*, which is a dimensional, dynamic entity, has played no role thus far.

Technically, the theory, as Einstein never stopped lamenting, wasn't even a mechanics, insofar as it did not offer a representation of individual quantum processes, or for that matter of anything else. As noted earlier, "observables," for the corresponding operators, and "states," for Hilbert-space vectors (in the language we use now), are other such terms: we never observe these "observables" or "states," or physically assign them to quantum objects (or to anything else), but only use them to predict, probabilistically, what will be observed in measuring instruments. To make these predictions, one will need Planck's constant, $h$, which thus enters as part of this new relation between the data in question and the mathematics of the theory.

That in general his new variables did not commute, $PQ - QP \neq 0$, was, again, an especially novel feature of Heisenberg's model. This feature, which was an automatic mathematical consequence of his choice of his variables, proved to be momentous physically. Most famously, it came to represent in the formalism Heisenberg's uncertainty relations constraining certain simultaneous measurements, such as those of the momentum ($P$) and the coordinate ($Q$), associated with a given quantum object in the mathematical formalism of quantum mechanics and (correlatively) the complementary nature of such measurements in Bohr's sense, explained below. Although it took a bit longer to realize the deeper physical nature of the quantum-mechanical situation to which Heisenberg's new mechanics responded, the noncommutative character of quantum variables should not be surprising. J. Schwinger instructively commented on the subject, in part following Bohr, whose influence is manifested in his lecture on quantum theory that I am about to cite. Schwinger does note the most commonly stated physical feature corresponding to this character, namely, that if one measures two physical properties in one order, and then in the other, the outcome would in general be different in quantum physics (while they are the same in classical physics). But he goes further in explaining why this is the case and what are its implications for the mathematical formalism of quantum theory:

> If we once recognize that the act of measurement introduces in the [microscopic] object of measurement changes which are not arbitrarily small, and which cannot be precisely controlled … then every time we make a measurement, we introduced a new physical situation and we can no longer be sure that the new physical situation corresponds to the same physical properties which we had obtained by an earlier measurement. In other words, if you measure two physical properties in one order, and then the other, which classically would absolutely make no difference, these in the microscopic realm are simply two different experiments …
>
> So, therefore, the mathematical scheme can certainly not be the assignment, the association, or the representation of physical properties by numbers because numbers do not have this property of depending upon the order in which the measurements are carried out. … We must instead look for a new mathematical scheme in which the order of performance of physical operations is represented by an order of performance of mathematical operations. (pp. 40-42, cited in Schweber 1994, p. 361)[41]

As must be clear from the preceding discussion, this is not how Heisenberg discovered the mathematical scheme of QM, in particular given that the noncommutativity of some among the operators representing quantum observables was not his starting point but a consequence of the multiplication rule for his matrices. The type of thinking described by Schwinger is more in accord with recent quantum-informational approaches of deriving the structure of quantum theory, primarily QTFD, from the (formalized) structure of quantum measurements.[42] The difficulty here is that any such scheme requires a great deal of nontrivial mathematics, especially at the time, such as that of Hilbert spaces (infinite-dimensional ones for continuous variables) over complex numbers and the use of operators to represent quantum variables, and, in addition, rules, such as Born's rule, by means of which the formalism is related to the probabilities or statistics of quantum predictions. It is not a matter of a *direct representation* of the order-dependent nature or structure of measurement in question by the formalism, in part because of the probabilistic or statistical nature of quantum predictions. We are so familiar with the relationships between the noncommutative nature of QM and the order in which some among quantum measurements are performed and the corresponding predictions are obtained, that we forget the *indirect* or, in Bohr's and Schwinger's terms, "symbolic," rather than directly representative, nature of these relationships.

It is worth adding that, as explained earlier and as Schwinger stresses in his lecture, no identical assignment of the single quantity is ever possible, or in any event ever guaranteed, in two "identically" prepared experiments in the way it can be in classical physics (Schweber 1994, p. 360). This is because quantum experiments cannot be controlled so as to identically prepare quantum objects but only so as to identically prepare the measuring instruments involved, because this behavior can be considered classical. The quantum strata of measuring instruments, through which they interact with quantum objects, do not affect these preparations but only the outcomes of actual measurements. On the other hand, this interaction itself is uncontrollable. This fact is central to Bohr's argument, which invokes this "finite and incontrollable interaction" at key junctures of his reply to EPR's paper (Bohr 1935, pp. 697, 700). It follows that the outcomes of repeated, identically prepared experiments, including those involving sequences of measurements, cannot be controlled even ideally (as they can be in classical physics), and these outcomes will, in general, be different. This circumstance makes statistical considerations unavoidable in any quantum experiment, and is reflected, among other things, in the statistical character of the uncertainty relations. The noncommutative nature of the corresponding quantum variables responds to this character as well, along with the uncertainty relations themselves. Heisenberg's famous formula, $\Delta q \Delta p \cong h$ (where $q$ is the coordinate, $p$ is the momentum in the corresponding direction), is, accordingly, statistical.

---

[41] My citation of the pages of the transcript of Schwinger's lecture, which is unpublished, follows S. Schweber, who cites the lecture at length in (Schweber 1994). "Microscopic" refers here to the *microscopic constitution* of quantum objects, and not to their actual size. Quantum objects could be macroscopic in size (Bose-Einstein condensates and SQUIDs are notable examples). Their quantum behavior, however, is determined by their microscopic quantum constitution, and thus, by the role of Planck's constant, $h$, in any quantum measurement used in observing them. Accordingly, one needs measuring instruments, described classically but capable of detecting quantum behavior, to observe macroscopic quantum objects as quantum.

[42] Schwinger was among those who anticipated these programs, specifically in (Schwinger 2001), which is, philosophically, in accord with his lecture cited here but offers a more rigorous approach to deriving QM from, in the words of Schwinger's subtitle, the "symbolism of atomic measurement." See (Jaeger 2016), for a helpful discussion of Schwinger's approach.

As Schweber says, in commenting on Schwinger's passage cited above, "Schwinger thus made plausible why in quantum mechanics physical properties are set in correspondence with noncommutative operators" (Schweber 1994, p. 361). It should be noted, however, that at least in Bohr's or the present view, and it appears, in Schwinger's view as well, the term "correspondence" should not be understood in the sense of the (realist) mathematical representation. For, at least in the present view, the operators in question only enable (along with other mathematical elements and structures of Hilbert spaces, and Born's or related rules added to them), the statistical predictions concerning properties observed in measuring instruments in quantum experiments, and do not correspond to any properties of quantum objects themselves. Schwinger appears to follow Bohr on this point, and he refers to Bohr and complementarity shortly following this passage. There are differences between Schwinger's and Bohr's interpretation overall, but they do not affect the main point at the moment, which concerns the relationships between the irreducibly statistical nature of quantum predictions and the noncommutativity of QM. I shall return to the question of the correspondence between mathematical and physical properties in quantum theory below in the context of quantum information theory.

The quantum-mechanical situation that emerged with Heisenberg's discovery of quantum mechanics and then Bohr's interpretation of it was eventually (sometime in the late 1930s) recast by Bohr in terms of his concept of "phenomenon," defined by what is observed in measuring instruments under the impact of quantum objects, in contradistinction to quantum objects themselves, which could not be observed or represented, or in the present view, even conceived of. According to Bohr:

> I advocated the application of the word phenomenon exclusively to refer to the *observations* obtained under specified circumstances, including an account of the whole experimental arrangement. In such terminology, the observational problem is free of any special intricacy since, in actual experiments, all observations are expressed by unambiguous statements referring, for instance, to the registration of the point at which an electron arrives at a photographic plate. Moreover, speaking in such a way is just suited to emphasize that the appropriate physical interpretation of the symbolic quantum-mechanical formalism amounts only to predictions, of determinate or statistical character, pertaining to individual phenomena appearing under conditions defined by classical physical concepts [describing the observable parts of measuring instruments]. (Bohr 1987, v. 2, p. 64)

Phenomena are irreducibly discrete in relation to each other, and, in Bohr's scheme, one cannot assume that there are continuous processes that connect them, especially classically causally, even in dealing with elemental individual processes and events. In nonrealist, RWR-view-based, interpretations, such as the one adopted by Bohr and implied in this passage, QM, again, only estimates the probabilities or statistics of the outcomes of discrete future events and tells us nothing about what happens between them. The formalism does not describe the data observed and hence quantum phenomena either. These data, defined by the bits of classical information, as part of the observed behavior of measuring instruments, are described by classical physics, which, however, cannot predict these data, either individually or collectively, including quantum correlations.

Part of Bohr's concept of phenomenon and the main reason for its introduction was that this concept "*in principle* exclude[s]" any representation or analysis, even if not a possible conception, of quantum objects and their behavior, at least, by means of QM (Bohr 1987, v. 2, p. 62). The concept of phenomenon is, thus, correlative to the RWR-type view, reached by Bohr at this stage of his thinking, keeping in mind possible differences between his and the present view, which places quantum objects and processes beyond the reach of human thought altogether. Physical quantities obtained in quantum measurements and defining the physical behavior of certain (classically described) parts of measuring instruments are *effects* of the interactions between quantum objects and these instruments. But these properties are no longer assumed to correspond to any properties pertaining to quantum objects themselves, even single such properties considered independently, rather than only certain joint properties, in accordance with the uncertainty relations. Bohr's earlier views allowed for this type of attribution *at the time of measurement* and only then. However, even this less radical view implied that the physical state of an object cannot be defined on the model of classical physics. This is because the latter model requires an unambiguous determination of both conjugate quantities for a given object at any moment of time and independently of measurement, which is not possible in quantum physics because of the uncertainty relations. In Bohr's ultimate view, however, which his concept of phenomenon reflects, an attribution *even of a single property* to any quantum object as such is *never possible—before, during, or after measurement*. The conditions that experimentally obtain in quantum experiments only allow one to rigorously specify measurable quantities that could physically pertain to measuring instruments. Even when we do not want to know the momentum or energy of a given quantum object and thus need not worry about the uncertainty relations, neither the exact *position* of this object itself nor the actual time at which this "position" is established is ever available and, hence, in any way verifiable. Any possible information concerning quantum objects as independent entities is lost in "the finite [quantum] and uncontrollable interaction" between quantum objects and measuring instruments (Bohr 1935, pp. 697, 700). However, this interaction leaves a mark in measuring instruments, a mark, a bit of information, that can be treated as a part of a permanent, objective record, which can be discussed, communicated, and so forth. The uncertainty relations remain valid, of course. But they now apply to the corresponding (classical) variables of suitably prepared measuring instruments, impacted by quantum objects. We can either prepare our instruments so as to measure or predict a change of momentum of certain parts of those instruments or so as to locate the spot that registers an impact by a quantum object, but never do both in the same experiment. The uncertainty relations are correlative to the complementary nature of these arrangements.

Bohr's argument implies and indeed arises from the assumption of the irreducible difference, reaching beyond Kant, between quantum phenomena and quantum objects:

> This necessity of discriminating in each experimental arrangement between those parts of the physical system considered which are to be treated as measuring instruments and those which constitute the objects under investigation may indeed be said to form a *principal distinction between classical and quantum-mechanical description of physical phenomena*. It is true that the place within each measuring procedure where this discrimination is made is in both cases largely a matter of convenience. While, however, in classical physics the distinction between object and measuring agencies does not entail any difference in the character of the description of the phenomena concerned, its fundamental importance in quantum theory, as we have seen, has its root in the indispensable use of classical concepts in the interpretation of all proper measurements, even though the classical theories do not suffice in accounting for the new types of regularities with which we are concerned in atomic physics. (Bohr 1935, p. 701, 697-697n)

This statement might suggest, and has suggested to some, that, while observable parts of measuring instruments are described by means of classical physics, the independent behavior of quantum objects is described by means of the quantum-mechanical formalism, which possibility, as explained, need not imply a phenomenal representation or visualization of these processes, only a mathematical representation of them by this formalism. This, however, is not the case. Bohr does say here that observable parts of measuring instruments are described by means of classical physics, again, with a crucial qualification that this description only concerns these observable parts, because measuring instruments also have quantum strata, through which they interact with quantum objects. But he does not say and does not mean (there is no evidence to conclude otherwise) that the independent behavior of quantum objects is represented by means of the quantum-mechanical formalism. As is clear, for example, from his description of his concept of phenomena cited above, this formalism is assumed by Bohr to have a strictly probabilistically or statistically predictive role, while, as explained in detail earlier, what "happens" between experiments cannot be represented, conceptually or possibly (and in the present, even if not Bohr's, view actually) even conceived of, including in terms of such concepts as "happening" or "occurrence." Nor can it be represented mathematically, including algebraically. The algebra of QM only predicts what can happens in experiments, although, as I said, in his later thinking, Heisenberg appears to give algebra (it is still algebra!) a chance to represent the ultimate constitution of nature.

While "it is true that the place within each measuring procedure where this discrimination [between the object and the measuring instrument] is made is … *largely* a matter of convenience" (emphasis added), it is true only largely, but not completely. As Bohr says: "In fact, it is an obvious consequence of [Bohr's] argumentation that in each experimental arrangement and measuring procedure we have only a free choice of this place within a region where the quantum-mechanical description of the process concerned is effectively equivalent with the classical description," in accordance with the correspondence principle (Bohr 1935, p. 701). In other words, quantum objects are always on the other side of the "cut" (as it became known) and may even be defined accordingly. At one end, by virtue of their classical nature, the individual effects observed in quantum experiments can be isolated materially and phenomenally (in the usual sense)—we can perceive and analyze them as such—once an experiment is performed. They cannot be separated from the process of their physical emergence by our even conceiving of, let alone analyzing, this process. This impossibility defines what Bohr sees as the indivisible wholeness of phenomena. By contrast, at the other end, quantum objects and processes can never be isolated, materially or phenomenally (in the usual sense of phenomenal as referring to what appears to our thought), because one cannot, even in principle, represent or even conceive of what *actually happens* at that level. Indeed, as discussed earlier, the concept of "actual" (or of "happening") only applies at the level of measuring instruments (Heisenberg 1962, pp. 47, 145).

The basis for Bohr's interpretation (in any of its versions) of quantum phenomena and QM was, I contend, Heisenberg's derivation of QM from the principles stated at the outset of this section, to which Bohr added the complementarity principle. Although Heisenberg's creativity and inventiveness were remarkable, his derivation was, arguably, not a strictly rigorous derivation, although it may depend on how one defines a "rigorous derivation." While borrowing the *form* of equations from classical mechanics by the mathematical correspondence principle was a logical deduction concerning part of the mathematical structure of QM, Heisenberg more "guessed" than derived the variables that he needed. The mathematical expression of the principles in question was only partially worked out and sometimes more intuited than properly developed, which took place a bit later in the work of Born, Jordan, and Heisenberg himself, and in some respects only with von Neumann's recasting of QM into his rigorous form of Hilbert-space formalism. Even when offering these more rigorously developed mathematical structures, a derivation of the mathematical model, such as the one offered by Heisenberg in his Chicago lecture (Heisenberg 1930), might still not be a rigorous derivation of the formalism from first principles. As he himself stated there:

> It should be distinctly understood, however, this [the deduction of the fundamental equation of quantum mechanics] cannot be a deduction in the mathematical sense of the word, since the equations to be obtained form themselves the *postulates* of the theory. Although made highly plausible by the following considerations [the same that led him to his discovery of QM], their ultimate justification lies in the agreement of their predictions with the experiment. (Heisenberg 1930, p. 108)

One might especially argue that it is not sufficiently first-principle-like to see the equations of QM as postulates and, in general, prefer a less mixed derivation of quantum theory than that of QM by Heisenberg or Schrödinger. One might, accordingly, envision a different type of derivation of QM from fundamental principles, especially by avoiding making the equations of quantum mechanics postulates of the theory, formally borrowed from classical physics, while Heisenberg's new variables were in fact a guess. Most recent work in this direction has been in quantum information theory in dealing with discrete variables and finite-dimensional Hilbert spaces (QTFD), as opposed to the infinite-dimensional Hilbert spaces needed for continuous variables. This work, however, has marked affinities with that of Heisenberg, which, as I argue here, in effect exhibits a quantum-informational, as well as and correlatively, structural-nonrealist thinking. I now turn to a discussion of some of this work where these affinities and certain key elements of structural realism are especially manifested, that of D'Ariano and coworkers and that of L. Hardy.

## 5. Structural Nonrealism and Quantum Information Theory

D'Ariano, Chiribella, and Perinotti's (DACP's) program, developed over the last decade and presented \comprehensively in their recent book (D'Ariano et al 2017), belongs to a particular trend in quantum information theory, and as most of the work along these lines, it deals with discrete variables and the corresponding (finite-dimensional) Hilbert spaces.[43]

---

[43] Among the key earlier works are J. A. Wheeler's "it-from-bit" manifesto, cited earlier (Wheeler 1990), (Zeilinger 1999), (Hardy 2001), and C. Fuchs's work (e.g., Fuchs 2003). Fuchs's work eventually "mutated" to QBism, a related but different program (e.g., Fuchs et al 2014). As mentioned above (note 40) a remarkable earlier precursor is J. Schwinger's approach to quantum

A rigorous (or at least more rigorous than that of founding figures, who discovered QM and QFT) derivation of QM, let alone QFT, from fundamental principles remains an open question. This is in part because one has to deal with continuous variables, where the technical application of the principles of quantum information theory, such as those used by DACP, is more complex. DACP's program does, however, have significant affinities with Bohr's and Heisenberg's thinking as discussed here. These affinities and, correlatively, the connections between DACP's program and structural nonrealism are my main interest in this section.

DACP's project is motivated by "a need for a deeper understanding of quantum theory [QTFD] in terms of fundamental principles," and by the aim of deriving QTFD from such principles, which, the authors contend, has never been quite achieved by their predecessors.[44] As indicated earlier, the fluctuations of terms, such as principles, axioms, postulates, used in this context may sometimes be confusing and obscure this common aim. DACP use the term "axioms" as well, referring to certain propositions on which their "principles" are based. On the other hand, while one can surmise their understanding of the term "principle" from their use of it, they do not define the concept of "principle" either, which is, as I noted, not uncommon. As earlier, I adopt the concept of principle, defined above, via Einstein's concept of a principle theory. This concept is, I would argue, in accord with DACP's use of principles, and their derivation of QTFD, as is that of Hardy, may be seen as that of a principle theory in Einstein's sense.

The main new feature of their approach is adding to the consideration of QTFD as an extension of probability theory (a view found in the works of their predecessors and, it might be added applicable to QM as well) "the crucial ingredient of *connectivity* among events" by using the operational framework of "circuits" (pp. 4-5). The framework of circuits has been similarly used by others as well, specifically by Hardy, as discussed below. This addition allows DACP "to derive key results of quantum information theory and general features of quantum theory [QTFD]" without first assuming Hilbert spaces. As explained earlier, unlike von Neumann, to whom DACP expressly refer for a contrast with their approach, Heisenberg, in his derivation of QM, did not begin with a mathematical

---

foundations (Schwinger 2001). Admittedly, there are other approaches to relating quantum foundations and quantum information, some of which, such as that advanced in (Grinbaum 2015), could have been considered in the present context, but most of this work, including that of A. Grinbaum, proceeds along different gradients from the one followed here.

[44] I put aside the question to what degree this derivation amounts to a fully rigorous derivation, which and, in the first place, the question of what could be considered as a fully rigorous derivation, would require a separate analysis. My main concern here is the relationship between their program and structural nonrealism, without, again, claiming that the latter is the authors' own view. Indeed, in his recent articles, D'Ariano takes a view that is closer to structural realism than structural nonrealism (e.g., D'Ariano 2017).

formalism either. He *arrived* at his formalism (essentially equivalent to the Hilbert-space one) from fundamental principles, even if, again, not fully rigorously *derived* it from these principles.

Among the principles adopted by DACP, the purification principle plays a uniquely important role as an essentially quantum principle, because conforming to it distinguishes QTFD from classical information theories. Notably, it is the single principle necessary to do so, which may not be surprising given the history of quantum theory and attempts at its principle or axiomatic derivations, beginning with that of von Neumann. Hardy's pioneering derivation similarly needed only one axiom, the continuity axiom, to do so as well (Hardy 2001). In nontechnical terms, the purification principle states that "every random preparation of a system can be achieved by a pure preparation of the system with an environment, in a way that is essentially unique" (D'Ariano et al 2017, p. 6). The principle originates in Schrödinger's insight in his response, in several papers, including his cat-paradox paper (Schrödinger 1935a), to the EPR paper (Einstein et al 1935). This insight led Schrödinger to his famous concept of entanglement. According to DACP:

> The purification principle stipulates that, whenever you are ignorant about the state of a system A, you can always claim that your ignorance comes from the fact that A is part of a large [composite] system AB, of which you have full knowledge. When you do this, the pure state that you have to assign to the composite system AB is determined by the state of A in an essentially unique way.
>
> The purification of mixed states is a peculiar feature—surely, not one that we experience in our everyday life. How can you claim that you know A *and* B if you don't have A alone? This counterintuitive feature has been noted in the early days of quantum theory, when Erwin Schrödinger famously wrote: "Another way of expressing the peculiar situation is: *the best possible knowledge of a whole does not necessarily include the best possible knowledge of all its parts.*" And, in the same paper: "I would not call that *one* but rather *the* characteristic trait of quantum mechanics, the one that enforced its entire departure from classical lines of thought" [Schrödinger 1935b, p. 555].
>
> This is a bold statement, if you think that it was made in 1935! Nowadays, however, there is plenty of evidence supporting it. Here are three arguments: first, we know that purification, combined with five [other] rather basic principles, picks up quantum theory among all possible theories one can imagine. Second, physicists have been trying to fabricate toy theories that exhibit quantum-like features without being quantum theory. While they succeeded with many features, up to now purification resists: if you want to purify mixed states, then quantum theory seems to be the only option, up to minimal variations. Third, many important features of quantum theory can be derived *directly* from purification, without the need of deriving quantum theory first. This fact strongly suggests that, when it comes to isolating what is specific of [to?] quantum theory, purification just hits the spot.
>
> The purification of mixed states is specifically quantum. But why should we assume it as a fundamental principle of Nature? At first, it looks like a weird feature—and it must

> look so, because quantum theory itself is weird and if you squeeze it inside a principle, it is likely that the principle looks weird too. However, on second thought one realizes that purification is a fundamental requirement: essentially, it is the link between physics and information theory. Information theory would not make sense without the notions of probability and mixed state, for the whole point about information is that there are things that we do not know in advance. But in the world of classical physics of Newton and Laplace, every event is determined and there is no space for information at the fundamental level. In principle, it does not make sense to toss a coin or to play a game of chance, for the outcome is already determined and, with sufficient technology and computational power, can always be predicted. In contrast, purification tells us that "ignorance is physical." Every mixed state can be generated in a single shot by a reliable procedure, which consists of putting two systems in a pure state and discarding one of them. As a result of this procedure, the remaining system will be a physical token of our ignorance. This discussion suggests that, only if purification holds, information can aspire to a fundamental role in physics. (D'Ariano et al 2017, pp. 168-169)

This elaboration needs, in my view, to be made more precise in discriminating between the physical states of a quantum system and the mathematical concept of quantum state as a vector, a state vector, in the Hilbert-space formalism, even if one take a more realism view of the mathematics concept of quantum state, as against a nonrealist, RWR-type, view, in which it is only a tool in making an expectation-catalog concerning the future behavior of the system considered, technically, again, concerning this system's interactions with the corresponding measuring instruments or, in DACP's language, circuits. DACP's formulation of the purification principle itself and their derivation of QTFD, which, again, does not assume Hilbert spaces first, allows one to differentiate these concepts. I would also argue that their overall scheme allows for a nonrealist interpretation, in part because of the role of the concept of circuit there, and I shall interpret it in this way here, without, however, claiming that this necessarily corresponds to DACP's own view. Also, the purification of mixed state is, technically, the fundamental principle of *our interactions with nature* by means of our experimental technology, rather than of nature itself, except of course insofar as we and our technologies are also nature. But then, again, there is no other principle of nature than those defined by us in our interactions with nature.

The main point at the moment is our quantum-informational knowledge, as always probabilistic in nature, depends on our informational ignorance, insofar as this knowledge cannot be complete in the classical sense, including, crucially, in dealing with individual quantum processes and events, which can be treated ideally deterministically in classical mechanics. In classical statistical physics, we do have an *analogous* incompleteness, which, however, only precludes such a treatment *in practice* and *not* in *principle*. As explained earlier, an ultimate underlying classical causality is assumed in considering classical physical systems, the complexity of which prevents us from making deterministic predictions concerning them, while, at the same time, the individual constituents of such systems could in principle, and sometimes even in practice, be treated deterministically. Accordingly, using the term "analogous" here requires caution. In quantum physics, "ignorance is physical," as DACP say, in that there is no continuity or classical causality, but, to return to Heisenberg's starting point in his derivation of QM, only discrete phenomena, defined by the interactions between quantum objects and the word, and transitions probabilities between them.

The purification principle could be related to Bohr's complementarity, which can be seen as a principle as well, or in any event, it may well be an effect of the same ingredient in the quantum constitution of nature that is responsible from complementarity or the EPR-type of effects, as discussed earlier. Bohr saw the EPR experiment and, thus, implicitly entanglement as a manifestation, perhaps the deepest manifestation, of complementarity, which already redefined the relationships between the whole and its parts, as against classical physics and relativity, or indeed all preceding conceptions of these relationships (Bohr 1935; Bohr 1987, v. 2, p. 59; Plotnitsky 2016, pp. 107-154). The mutual exclusivity of "parts" that define complementarity also means that these parts are never parts of the same whole. Each part is the only whole there is at any given point. Nor can an entangled system be seen as merely a sum of its parts, in the way parts sum up to a whole in classical physics or relativity. One must keep in mind a) that entanglement represents a particular, if extremely and even uniquely important, case of complementarity, which is a more general concept (e.g., Plotnitsky 2016, pp. 136-155); and b) that entangled states, at least as understood here, are part of the mathematical technology of our statistical predictions concerning the outcomes of quantum experiments, rather than represent any real quantum-level properties. It may, however, be said that entanglement and purification represent a completely different type of relation between physical systems because all possible measurements or predictions concerning them are defined by the *complementary* nature of the variables involved in these measurements or predictions.

Thinking in terms of "circuits" is close to Bohr's thinking concerning the role of measuring instruments in the constitution of quantum phenomena, as distinguished from quantum objects, which give rise to quantum phenomena by interacting with measuring instruments but which are never observable. In Bohr's interpretation, the concept of complementarity applies strictly to certain specifiable measuring arrangements. Circuits and their arrangements, too, embody the technological structure of measuring instruments capable of detecting and measuring quantum events, and also enabling the probabilistic predictions of future events. Their arrangements and operations are enabled by rules that should ideally be derived from certain sufficiently natural assumptions. Circuits and their arrangements are described classically, and thus physically embody the structure of quantum information as a particular structure, a particular form of organization, of classical information, which can be used, as in DACP's (or Hardy's) program, to *derive* the mathematical formalism of QTFD, comprised of the set of *mathematical structures*, which essentially amount to those of Hilbert spaces over complex numbers. Bohr (who was also dealing primarily with continuous rather than discrete variables) was not concerned with such a derivation, but only with an interpretation of an already established formalism.

While, however, it has multiple historical connections and resonances, such as those just indicated, the purification principle has additional features to those provided by Schrödinger's elaboration, cited above, concerning "*the* characteristic trait of quantum mechanics." Thus, Schrödinger's analysis does not contain the requirement that if two pure states of a composite system AB have the same marginal on system A, then they are connected by some reversible transformation on system B, which amounts to the assumption that all purifications of a given mixed state are equivalent under local reversible operations (D'Ariano et al 2017, pp. 169-170). Schrödinger never saw this "trait" as a fundamental principle either: it was only a "trait," and moreover, "*the* characteristic trait of *quantum mechanics*," which he saw as "a doctrine born of distress," and not necessarily a fundamental feature of *nature* (Schrödinger 1935a, p. 154). By contrast, the purification principle is viewed by DACP "as a fundamental principle of Nature," even if with the qualification given above concerning the role of our interactions with nature is defining any principle of nature (D'Ariano et al 2017, p. 169).

While indispensable in the authors' operational derivation of QTFD, the purification principle in itself is not, and cannot be, sufficient to do so. They need five additional postulates:

> (1) *Causality* [essentially locality]. Measurement results cannot depend on what is done on the system at the output of measurements. Equivalently: no signal can be sent from the future to the past.
> (2) *Local discriminability*. We can reconstruct the joint state of multiple systems by performing local measurements on each system.
> (3) *Perfect distinguishability*. Every state that is not completely mixed [i.e., if it cannot be obtained as a mixture from any other state] can be perfectly distinguished from some other state.
> (4) *Ideal compression*. Every source of information can be encoded in a lossless and maximally efficient fashion (*lossless* means perfectly decodable, *maximally efficient* means that every state of the encoding system represents a state in the [information] source).
> (5) *Atomicity of Composition*. No side information can hide the composition of two atomic transformations. Equivalently: the sequential composition of two precisely known transformations is precisely known. (D'Ariano et al 2017, p. 6)

These five informational postulates define a large class of classical probabilistic informational theories, while one, again, needs one and only one additional postulate giving rise to the corresponding principle, the purification principle, to distinguish QTFD in this case. The appearance of these additional principles is not surprising. Heisenberg's grounding principles, the quantum discreteness, QD, principle and the quantum probability or statistics, QP/QS, principles, were not sufficient for him to derive, or again, to arrive at QM, either. To do so, he needed the correspondence principle, to which he gave a rigorous mathematical form, requiring that both the equation and variables used convert into those of classical mechanics in the classical limit, which gave him, as it were, half of the architecture of quantum theory. The other half, essentially amounting to a Hilbert-space architecture, was supplied by his new matrix variables, which were more guessed than rigorously derived by Heisenberg from first principles. There is of course a difference, because, unlike those of DACP's, all of Heisenberg's principles were essentially quantum-mechanical in nature and did not apply to classical physics, apart from the fact that the equations used were formally those of classical mechanics.

There are, however, instructive parallels between DACP's and Heisenberg's approaches. The QP/QS principle is present in both cases, given that DACP see QTFD as an operational-probabilistic theory of a special type, defined by the purification postulate. As they say in their earlier article: "The operational-probabilistic framework combines the operational language of circuits with the toolbox of probability theory: on the one hand experiments are described by circuits resulting from the connection of physical devices, on the other hand each device in the circuit can have classical outcomes and the theory provides the probability distribution of outcomes when the devices are connected to form closed circuits (that is, circuits that start with a preparation and end with a measurement)" (Chiribella et al 2011, p. 3). This is similar to Heisenberg's thinking in his paper introducing QM. It is true that the concept of "circuit" in not found in Heisenberg and its role in DACP's approach is, again, ultimately closer to Bohr's view of the role of measuring apparatuses and his concept of phenomenon, defined by this role. As I argue here, however, the idea that in quantum theory we only deal with transition probabilities between the outcomes of the interactions between quantum objects and measuring instruments was introduced by Heisenberg, along with and defining quantum mechanics, and was then adopted and developed by Bohr in his interpretation. Heisenberg found the formalism of QM by using the mathematical correspondence principle, which allowed him to adopt the equations of classical mechanics, while changing the classical variables, the combination that defined the structure of QM. Heisenberg needed new variables for these formally classical equations because the classical variables did not give Bohr's frequencies rules for spectra, and the corresponding probabilities or statistics. With the help of profound physical and mathematical intuitions, added to more rigorous physical and mathematical argumentation, Heisenberg discovered that these rules are satisfied by, in general, noncommuting matrix variables with complex coefficients, related to amplitudes, from which one derives, by means of a Born-type rule, probabilities or statistics for transitions between stationary states, manifested in spectra observed in measuring devices. It is in this way that Heisenberg's derivation was essentially related to the fact that measuring instruments are devices with classically describable observable parts, which are, thus, akin to "operational circuits."

DACP's approach, again, arises from their aim to arrive at the mathematical structure of QTFD in a more first-principle-like way, in particular, independently of classical physics, which, because of the correspondence principle was central to Heisenberg. (Classical physics, to begin with, does not have discrete variables, such as spin, which are purely quantum and with which QTFD is associated.) This is accomplished by using the rules governing the structure of operational devices, circuits. These rules are *more empirical*, but they are *not completely empirical*, because circuits *are* given a mathematical structure, even though this structure may be partially defined by the organization, structure, of the experimental arrangement in which circuits appear.[45] But then, as noted one several occasions in this article, nothing is ever completely empirical even in observation. This move of DACP's is *parallel* to Heisenberg's arrangement of the quantities he used into the square tables, matrices, which was a mathematical invention giving structure to the manifolds of physical quantities, ultimately linked to probabilities of transitions between stationary states. In fact, this could even be seen in relation to circuits where these quantities are observed or (probabilistically) predicted as spectra. Heisenberg, again, needed other elements to establish this architecture fully, admittedly, in a way quite different from how DACP aim to do this. Nevertheless, something akin the structure of circuits was essential for establishing the mathematical structure of Heisenberg's scheme. This scheme related to this structure only by providing the probabilities or statistics of the outcomes of discrete quantum experiments, in accord with the QD and QP/QS principles, without providing any representation of quantum processes themselves. In other words, in accordance with the principles of structural nonrealism, in both cases, DACP's and Heisenberg's, one deals with two structures: the mathematical structure of the theory used, determined by a (mathematical) law of structure, and the correlational informational structure of quantum phenomena (observed in circuits), for the emergence of which the theory provides no law. These two structures are only related, and given the nature of the second structure, could only be related, by means of statistical predictions. The mathematized structure of circuits mediates these relationships. Deriving QTFD from the principles proposed by DACP requires enormous technical work, which should come as no surprise, as the derivation of QM was far from easy for Heisenberg and other founding figures as well, a fact often forgotten, in part, ironically, by virtue of the power and effectiveness of the mathematics of QM. It is next to impossible to do this derivation justice here. I would like, however, to consider further the concept of circuit, still focusing on its foundational rather than quantum-computational aspects (the concept of circuit originated in quantum computing).

---

[45] Hence, as I indicated earlier (note 44), to what degree the formalism of QTFD is ultimately *derived* within DACP's scheme or, again, what constitutes such a derivation requires much additional consideration, which cannot be pursued here. One might even question, and some did, the necessity of such a "fully rigorous definition." After all, Heisenberg, at least, as his scheme was developed by Born and Jordan, and then differently Dirac (inspired by Heisenberg as well) did establish a correct theory, and then Dirac similarly discovered QED as well. As Hardy suggested, at least by his title, in his pioneering paper (Hardy 2001), it may be more a matter what are "reasonable" initial postulates, although the "reasonableness" of such postulates is not a simple or unconditional matter either. It goes without saying that these qualifications in no way diminish the significance of DACP's or Hardy's work, or that of others pursuing these lines of thinking and research.

As explained above, circuits represent the arrangements of measuring instruments that are capable of quantum measurements and predictions, which are probabilistic or statistical, and sometimes, as with the EPR or the EPR-Bohm type of experiments are correlated. A realist representation of these arrangements is possible and unproblematic because they are described by classical physics, even though they interact with quantum objects, and thus have a quantum stratum, which enables this interaction, disregarded by this classical representation, luckily, without any detriment to either measurements or predictions. The specific structures and the properties of circuits help and ideally enable us to derive the formalism of QTFD, that is, the "arrangements" of elements (such as Hilbert-space operators) and thus the structure, governed by a mathematical law, that define this formalism. This formalism, again, has strictly probabilistically or statistically predictive relations to what is observed, without representing quantum objects and processes.

One can gain further insights into the structure of circuits from Hardy's work. Hardy arrives at a different set of main assumptions necessary to derive QTFD than those of DACP, but the main strategy is the same: establishing the architecture of circuits that, perhaps with additional axioms, would allow one to derive the mathematical formalism of QTFD. According to Hardy:

> Circuits have:
> • A setting, s(H), given by specifying the setting on each operation.
> • An outcome set, o(H), given by specifying the outcome set at each operation (equals o(A) × o(B) × o(C) × o(D) × o(E) in this case). We say the fragment "happened" if the outcome is in the outcome set.
> • A wiring, w (E), given by specifying which input/output pairs are wired together.
> (Hardy 2013, p. 7)

With this definition in hand, I shall comment on some of Hardy's fundamental assumptions as discussed by him in a different paper, which make my main point more transparent. Hardy proceeds as follows:

> We will make two assumptions to set up the framework in this paper. …
> **Assumption 1**. The probability, Prob (A), for any circuit, A (this has no open inputs or outputs), is well conditioned. It is therefore determined by the operations and the wiring of the circuit alone and is independent of settings and outcomes elsewhere. (Hardy 2010, p. 11)

This is a physical postulate, essentially that of spatial and temporal locality, combined with probability or statistics, along the lines of the QP/QS principle. The task now becomes how to derive a QTFD that could correctly predict these probabilities. One needs another assumption:

> **Assumption 2**: **Operations are fully decomposable**. …. In words we will say that any operation is equivalent to a linear combination of operations each of which consists of an effect for each input and a preparation for each output. …
>
> Assumption 2 introduces a *subtly* different attitude than the usual one concerning how we think about what an operation is. Usually we think of operations as effecting a transformation on systems as they pass through. Here we think of an operation as corresponding to a bunch of separate effects and preparations. We need not think of systems as things that preserve their identity as they pass through—we do not use the same labels for wires coming out as going in. This is certainly a more natural attitude when there can be different numbers of input and output systems and when they can be of different types. Both classical and quantum transformations satisfy this assumption. In spite of the different attitude just mentioned, we can implement arbitrary transformations, such as unitary transformations in quantum theory, by taking an appropriate sum over such effect and preparation operations. (Hardy 2010, pp. 19-20)

This rethinking of the concept of operation is important, especially if one adopts a nonrealist, RWR-type, view. For, consistently with this view, an "operation" is now defined in terms of observable "effects" of the interactions between quantum objects and measuring instruments (or circuits), and not in terms of what happens, even in the course of these interactions (let alone apart from them), to the quantum objects or systems, *considered as independent systems*. It is also true and useful (in allowing one to treat classical and quantum informational systems within the same overall informational framework) that we can treat classical systems in this way as well. In the classical case, however, we can also, equivalently, use a more conventional concept of operation mentioned here, which is not the case in quantum theory, at least, again, if one adopts a nonrealist view. This suggests that quantum systems or quantum operations must be distinguished from classical ones by means of certain additional principles or postulates, similarly to the way DACP use the purification principle, or Hardy used his continuity axiom in his earlier work (Hardy 2001). After a technical discussion of "duotensors" (which I put aside), Hardy suggests a principle:

> *Physics to mathematics correspondence principle*. For any physical theory, there [exist] a small number of simple hybrid statements that enable us to translate from the physical description to the corresponding mathematical calculation such that the mathematical calculation (in appropriate notation) looks the same as the physical description (in appropriate notation).

Such a principle might be useful in obtaining new physical theories (such as a theory of quantum gravity). Related ideas to this have been considered by category theorists. A category of physical processes can be defined corresponding to the physical description. A category corresponding to the mathematical calculation can also be given. The mapping from the first category to the second is given by a functor (this takes us from one category to another). (Hardy 2010, p. 39)

The language of correspondence should not mislead one into relating this principle to Bohr's correspondence principle, in either Bohr's initial form or in Heisenberg's mathematical form. The correspondence principle deals with the correspondence between different physical theories (such as classical mechanics and quantum mechanics, or later on between QM and QFT), possibly established mathematically, insofar as their predictions would coincide in the regions when both theories could be used, which converts QM into classical mechanics and QFT into QM in the corresponding limits. By contrast, Hardy's "hybrid" construction serves to give the structures to circuits so that these structures could then be "translated" into a proper formalism of QTFD, which can then be related to what is observed.

As explained earlier, circuits and their arrangements embody the technological structure of measuring instruments capable of detecting quantum events and enabling, by means of QTFD, the probabilistic or statistical predictions of future events, in other words, a structure that may be mathematizable and translatable into the mathematical formalism of QTFD (Hardy's "physics to mathematics correspondence principle"). The second category can be clearly defined mathematically, say, as that of Hilbert spaces and morphisms between them. On the other hand, the first category, that is, the structure of its objects, which are formalized circuits, and the morphisms between them, is a more complex matter, which would require a separate discussion. This discussion cannot be undertaken here, except for indicating the following important aspect of the situation.

As indicated earlier in considering both Schwinger's argument (concerning the relationships between the noncommutative nature of QM and the order-dependence quantum measurement) and DACP's approach, one need not, and in the present, RWR-type, view, should not, expect a direct representational correspondence between the (mathematized) structures of circuits and the mathematical structures of the formalism of QTFD. Instead, it is a matter of a translation, of defining a functor, between two categories of different structured objects—circuits and that of Hilbert spaces over complex numbers. In this respect, Hardy's formulation above "for any physical theory, there [exist] a small number of simple hybrid statements that enable us to translate from the physical description to the corresponding mathematical calculation such that the mathematical calculation (in appropriate notation) *looks the same* as the physical description (in appropriate notation)" (emphasis added) might require further qualification as concerns the meaning of the expression "looks the same." His characterization of the functor in question as "*virtually* direct" (emphasis added) may need further qualification as well. Similar questions may be posed concerning DACP's argumentation, but Hardy's use of category theory helps one to better approach these questions. In the view adopted here, the experimentally established structure of physical operations could be categorically formalized (the first category invoked by Hardy) and categorically or functorially mapped on the structure of certain mathematical operations belonging to mathematical objects, such as Hilbert spaces, which, again, form a well-defined category (the second category invoked by Hardy). These operations enable one to predict, in probabilistic or statistical terms, certain possible future data defined within the same experimentally established and categorically formalized structure. In other words, the suitably mathematized structures of circuits formalize the arrangements of measuring instruments involved and, as a result, enable one to derive the mathematical structure of QTFD, which enables correct probabilistic predictions concerning the data associated with these structures of circuits (on the basis of measurements equally associated with these structures). In essence, the mathematical structure of QTFD is analogous, and likely to be equivalent, to the Hilbert-space of the formalism. Arguably, however, this derivation is only possible with the help of certain additional fundamental principles, such as the purification principle of DACP. One always needs, for example, Born's rule or one or another equivalent rule.

Indeed, the essence of category theory consists in establishing the relationships between multiplicities, "categories," of objects of *different* types, for example and in particular, geometrical or, more generally, topological objects and algebraic objects, rather than, apart from trivial cases, directly (isomorphically) mapping the objects of the first category on those of the second. The concept of "category" was introduced in the field of algebraic topology and then extended to algebraic geometry in order to help to study certain algebraic invariants, such as groups, associated with topological spaces. In contradistinction to geometry, defined, as a mathematical discipline, by the concept of measurement (geo-*metry*), topology, as a mathematical discipline, is defined by associating an algebraic structure or a set of structures, most especially groups, such as homotopy or cohomology groups (which came to play an important role in QFT and string theory) to a topological space as a structured object. The structure of a topological space is defined by its continuities and discontinuities, and not by its geometry, even if it had a geometry, and not all topological spaces do. Insofar as one deforms a given figure continuously (i.e. insofar as one does not separate points previously connected and, conversely, does not connect points previously separated), the resulting figure or space is considered the same, which sameness is difficult and perhaps impossible to think by means of our general phenomenal intuition and, hence, is mathematical. The proper mathematical term is "topological equivalence." Thus, all spheres, of whatever size and however deformed, are topologically equivalent, despite the fact that some of the resulting objects are no longer spheres geometrically speaking. Such figures are, however, topologically distinct from tori, because spheres and tori cannot be converted into each other without disjoining their connected points or joining the disconnected ones: the holes in tori make this impossible. In modern (categorical) algebraic topology, mathematical objects of each type are arranged in "categories" and, within each category, related to one another by collections of mappings or "morphisms" (also known as "arrows"), and categories themselves are related by "functors." The category of topological spaces and their morphisms (or a subcategory, such as that of differential manifolds and their morphisms) becomes related to a category of algebraic objects, such as, again, groups, and their morphisms. These relations between these two categories, topological and algebraic, allow one to extract an enormous amount of information concerning topological spaces and sometimes, conversely, groups through topological spaces to which they may be associated.

The main point at the moment is the different, rather than isomorphic or homomorphic, nature of objects related, categorically, in terms of functors and not in terms of (realist-like) representations between them. There are morphisms, including isomorphism, in each category and functors between categories. Hardy's "physics to mathematics correspondence principle," becomes, if one adopts the view just outlined, becomes the "physics to mathematics *functoriality* principle:" that is, it may be seen in terms of this type of functorial relationship between the category of circuits (which category still requires a proper definition, especially as concerns its "morphisms") and the category of Hilbert spaces, finite-dimensional ones in the case of QTFD. These relationships enable one to use the "topological" structure of circuits (which provides the input data) and the algebraic structures of Hilbert spaces (functorially derived from this topology of circuits) to predict, in probabilistic or statistical terms, the data that is expected to be found in the same or possibly some other topological structure of circuits. I use the term "topology" metaphorically for the moment, while, however, keeping in mind that the structures of circuits may be given topology in a mathematically rigorous sense.

An important and difficult question is that of the relationship between the structure of the circuits, defined by the corresponding experimental arrangements, and the infinite-dimensional mathematical architecture of QM, or the same relationship in QFT. Consider the double-slit experiment, say, in the interference pattern setup. It is a circuit, which embodies (or could be represented by a scheme that embodies) preparations, measurements, and predictions, all manifested in the emergence of the interference pattern. I would not presume to be able to describe this setup as a circuit, let alone in any way "mathematize" it (although mathematizing in effect amounts to what it would mean to describe a circuit). But it is a circuit nevertheless, a complex one, albeit child's play in comparison to the circuitry found in high-energy quantum physics, such as that of the Large

Hadron Collider (LHC), which led to the detection of the Higgs boson.

Such questions will, however, need be addressed if one is to extend such programs to derivations of QTFD to QM or to QED and QFT, or beyond as both Hardy and DACP aspire to do, along somewhat different gradients. For, important as this task may be, it is hardly sufficient for these programs to merely derive already established theories. Their ultimate value lies in solving still outstanding problems, in what they can do for the future of fundamental physics. Hardy aims to rethink general relativity in operational terms, analogous to those of QTFD, and then to reach, in principle, quantum gravity by combining the operational frameworks of quantum theory (including QFT) and general relativity (Hardy 2007). D'Ariano and co-workers, by contrast, appear first to move from QTFD to QFT. In their more recent work, they aim to develop a new approach to QFT, which, as based on the concept of quantum cellular automata, is different from the operational framework discussed thus far, but shares with it certain key informational features and, most especially, the aim of developing quantum field theory from fundamental (first) principles. D'Ariano and Pernotti's derivation of Dirac's equation is a step in this direction, for now in the absence of an external field, essential to the proper QED (D'Ariano and Perinotti 2014; D'Ariano 2017). Unlike Dirac's own or other previous derivations of the equation, D'Ariano and Pernotti's derivation only uses, along with other principles (homogeneity, isotropy, and unitarity), the principle of locality, rather than those of special relativity. The approach may, in the authors' view, also offer new possibilities for fundamental physics on Planck's scale, thus suggesting a potential extension of quantum informational theory as far as physics can envision now.

## 6. Conclusion

Quantum information theory lends itself to the framework of structural nonrealism because, in the present interpretation of it, as extending from Heisenberg thinking in his discovery of quantum mechanics, it amplifies the nonrealist, RWR-type, view of quantum theory as defined by the probabilistic and statistical relationships between structures, specifically, between structures defined by laws of structure and structures that preclude us from establishing the laws of their emergence. There is no theoretical physics and no human thought apart from structures and laws. But the *emergence* of the structures defining quantum phenomena and quantum information is structureless and lawless. Indeed, the nature of this emergence might be beyond the reach of thought altogether.

**Acknowledgments.** I would like to thank Mauro D'Ariano, Laurent Freidel, Christopher A. Fuchs, Lucien Hardy, Gregg Jaeger, Andrei Khrennikov, and Paolo Perinotti for valuable discussions concerning the subjects addressed in this article.

# References

Aronson, S.: *Quantum Computing since Democritus*. Cambridge, UK: Cambridge University Press (2013)

Aspect, A., Dalibard, J., Roger, G.: Experimental test of Bell's inequalities using time varying analyzers, Physical Review Letters **49**, 1804-1807 (1982)

Badiou, A.: *Briefings on Existence* (tr. Madarasz, N.) Albany, NY: SUNY Press (2007)

Barbour, J. B.: *The end of time: The next revolution in physics*. Oxford, UK: Oxford University Press (1999)

Bell J.S.: *Speakable and unspeakable in quantum mechanics*. Cambridge, UK: Cambridge University Press (2004)

Bohr, N.: On the constitution of atoms and Molecules (Part 1), Philosophical Magazine **26** (151), 1-25 (1913)

Bohr, N.: Can Quantum-Mechanical Description of Physical Reality Be Considered Complete? Phys. Rev. 48, 696-702 (1935)

Bohr, N.: Causality and complementarity. In: Faye, J., Folse, H. J., (eds.) *The Philosophical Writings of Niels Bohr, Volume 4: Causality and Complementarity, Supplementary Papers*, 83-91. Ox Bow Press, Woodbridge, CT 1994 (1937)

Bohr, N: The Causality Problem in Atomic Physics. In: Faye, J., Folse, H.J. (eds.) *The Philosophical Writings of Niels Bohr, Volume 4: Causality and Complementarity, Supplementary Papers*, 94–121. Ox Bow Press, Woodbridge, CT, 1987 (1938)

Bohr, N.: *The Philosophical Writings of Niels Bohr*, 3 vols. Ox Bow Press, Woodbridge (1987)

Born, M., Jordan, P.: Zur Quantenmechanik, Z. für Physik **34**, 858 (1925)

Born, M., Heisenberg, W., Jordan, P.: On quantum mechanics. In: Van der Waerden, B. L. (ed), *Sources of Quantum Mechanics*, Dover; New York, 1968, 321–385 (1926)

Brunner N., Gühne O., Huber, M (eds): Special issue on 50 years of Bell's theorem. J. Phys. A **42**, 424024 (2014)

Butterfield, J., Isham, C. J: Spacetime and the philosophical challenge of quantum gravity. In: Callender, C., Huggett, N. (eds) *Physics Meets Philosophy at the Planck Scale: Contemporary Theories of Quantum Gravity*. Cambridge University Press; Cambridge 2001, 33–89 (2001)

Cassinelli, G., Lahti, P.: An Axiomatic Basis for Quantum Mechanics. Found. Phys. **46** (10), 1341-1373 (2016)

Chiribella, G., D'Ariano, G.M., Perinotti, P.: Informational derivation of quantum theory. Phys. Rev. A 84, 012311-1–012311-39 (2011)

Connes, A.: *Noncommutative Geometry* (tr. Berberian, S. K., ed. Rieffel, M. A.) San Diego, CA. Academic Press (1994)

Corry, L.: *Modern Algebra and the Rise of Mathematical Structures*. Boston Birkhäuser (2013)

Cushing J. T., McMullin E. (eds): *Philosophical consequences of quantum theory: reflections on Bell's theo*rem. Notre Dame, IN: Notre Dame University Press (1989)

D'Ariano, G. M.: Physics without Physics. Int. J. Theor Phys **56** (1), 97-38 (2017)

D'Ariano, G.M., Chiribella, G., and Perinotti, P.: *Quantum Theory from First Principles: An Informational Approach*. Cambridge University Press, Cambridge (2017)

D'Ariano, G.M., Perinotti, P.: Derivation of the Dirac equation from principles of information processing. Phys Rev A **90,** 062106 (2014)

Darigold, O. *From c-Numbers to q-Numbers: The Classical Analogy in the History of Quantum Theory*. University of California Press, Berkeley, CA (1993).

De Finetti, B.: *Philosophical Lectures on Probability* (tr. Hosny, H.) Springer, Berlin (2008)

Deleuze, G., Guattari, F.: *What is Philosophy?* (tr. Tomlinson, H., Burchell, G.) Columbia University Press, New York (1994)

Derrida, J.: *Writing and Difference* (tr. Bass, A.) Johns Hopkins University Press, Baltimore (1980)

Dirac, P. A. M.: *The Principles of Quantum Mechanics*, Clarendon, Oxford (1930)

Dirac, P. A. M.: *The Principles of Quantum Mechanics* (4th edition). Clarendon, Oxford, rpt. 1995 (1958)

Dzhafarov, E., Jordan, S., Zhang R., Cervantes V. (eds.): *Reality, contextuaity, and probability in quantum theory and beyond*, Singapore: World Scientific (2016)

Einstein, A.: Strahlungs-emission und -absorption nach der Quantentheorie. Deutsche Physikalische Gesellschaft Verhandlungen **18**, 318–323 (1916a)

Einstein, A.: Zur Quantentheorie der Strahlung. Physikalische Gesellschaft Zurich **18**, 173–177 (1916b)

Einstein, A.: What is the Theory of Relativity? (1919). In: Einstein, A. *Ideas and Opinions*, pp. 227-231, Bonanza Books, New York, 1954 (1919)

Einstein, A.: Physics and reality. Journal of the Franklin Institute, 221, 349–382 (1936)

Einstein, A.: Autobiographical notes (tr. Schillp, P. A.) La Salle, IL: Open Court (1949)

Einstein, A., Podolsky, B., and Rosen, N.: Can Quantum-Mechanical Description of Physical Reality be Considered Complete?, 138-141. In Wheeler, J.A., Zurek, W.H. (eds), *Quantum Theory and Measurement*, Princeton University Press, Princeton NJ, 1983 (1935)

Ellis J., Amati D. (eds): *Quantum reflections*. Cambridge, UK: Cambridge University Press (2000)

Feynman, R.: *QED: The strange theory of light and matter*. Princeton University Press, Princeton, NJ (1985)

Freidel, L.: On the discovery of quantum mechanics by Heisenberg, Born, and Jordan. (Unpublished) (2016)

Frigg, R., and Hartmann, S.: Models in Science, in Zalta, E. N., ed., The Stanford Encyclopedia of Philosophy (Fall 2012 Edition), URL = http://plato.stanford.edu/archives/fall2012/entries/models-science/ (2012)

Fuchs, C. A.: Quantum mechanics as quantum information, mostly. J. Mod. Opt. **50**, 987-223 (2003)

Fuchs, C. A., Mermin, N. D., Schack, R.: An introduction to QBism with an application to the locality of quantum mechanics. Am. J. Phys. **82**, 749. http://dx.doi. org/10.1119/1.4874855 (2014)

Griffiths, R. B.; Quantum Information: What's It All About? arXiv:1710.08520v2 [quantum-ph] 30 November 2017

Grinbaum, A.: Quantum Theory as a Critical Regime of Language Dynamics, Found. Phys. **45** (10): 1341-1350 (2015)

Hardy, L.: Quantum mechanics from five reasonable axioms. arXiv:quant-ph/0101012v4 (2001)

Hardy, L.: Towards quantum gravity: a framework for probabilistic theories with non-fixed causal structure," *J. Phys.* A **40**, 3081–3099 (2007)

Hardy, L.: A formalism-local framework for general probabilistic theories, including quantum theory, URL: <arXiv.1005.5164 [quant-ph]> (2010)

Hardy, L.: Foliable operational structures for general probabilistic theory. In: Halvorson, H. (ed.) Deep beauty: Understanding the Quantum World through Mathematical Innovation, pp. 409-442. Cambridge University Press, Cambridge, 2011 (2011)

Háyek, A.: Interpretation of Probability, Stanford Encyclopedia of Philosophy (Winter 2014 edition), E. N. Zalta (ed.) http://plato.stanford.edu/archives/win2012/entries/probability-interpret/ (2014)

Heisenberg, W.: Quantum-theoretical re-interpretation of kinematical and mechanical relations. In: Van der Waerden, B.L. (ed.) *Sources of Quantum Mechanics*, pp. 261–277. Dover, New York, Reprint 1968 (1925)

Heisenberg, W.: *The Physical Principles of the Quantum Theory* (tr. Eckhart, K., Hoyt, F.C.) Dover, New York, rpt. 1949 (1925)

Heisenberg, W.: *Physics and Philosophy: The Revolution in Modern Science,* Harper & Row, New York (1962)

Heisenberg, W.: *Encounters with Einstein, and other essays on people, places, and particles*, Princeton University Press, Princeton, NJ (1989)

Interpretations of quantum mechanics. *Wikipedia*. Interpretations of Quantum Mechanics. https://en.wikipedia.org/wiki/Interpretations_of_quantum_mechanics

Jaeger, G.: *Quantum Objects: Non-local Correlation, Causality and Objective Indefiniteness in the Quantum World*. Springer, New York (2013)

Jaeger, G.: Grounding the randomness of quantum measurement. Philos. Trans. Royal Soc. **A** . DOI: 10.1098/rsta.2015.0238 (2016)

Jaynes, E.T.: *Probability Theory: The Logic of Science*. Cambridge University Press, Cambridge (2003)

Kant I.: *Critique of pure reason* (tr. Guyer, P., Wood, A.W.). Cambridge, UK: Cambridge University Press (1997)

Khrennikov, A.: *Interpretations of Probability*. de Gruyter, Berlin (2009)

Khrennikov, A.: Quantum probabilities and violation of CHSH-inequality from classical random signals and threshold type detection scheme. Prog. Theor. Phys. **128**, 31–58. (doi:10.1143/PTP.128.31) (2012)

Ladyman, J.: Structural Realism. The Stanford Encyclopedia of Philosophy (Winter 2016 Edition), Edward N. Zalta (ed.), URL = https://plato.stanford.edu/archives/win2016/entries/structural-realism/ (2016)

Lucretius, T. C.: *On the nature of the Universe* (tr. Melville, R.), Oxford: Oxford University Press (2009)

Mehra, J., and Rechenberg, H.: *The Historical Development of Quantum Theory*, 6 vols. Springer, Berlin (2001)

Pauli, W.: Zur Quantenmechanik des magnetischen Elektrons, Z. Physik **43**, 601-625 (1927)

Plato. Phaedo. In: *The collected dialogues of Plato* (ed. Hamilton, E., Cairns, H.), 40–98, Princeton, NJ: Princeton University Press (2005)

Plotnitsky, A.: Quantum Atomicity and Quantum Information: Bohr, Heisenberg, and Quantum Mechanics as an Information Theory, in *Quantum Theory: Reexamination of Foundations 2,* ed. Andrei Khrennikov, Växjö, Sweden: Växjö University Press (2002)

Plotnitsky, A.: *Epistemology and Probability: Bohr, Heisenberg, Schrödinger and the Nature of Quantum-Theoretical Thinking*. Springer, New York (2009)

Plotnitsky, A.: *The Principles of Quantum Theory, from Planck's Quanta to the Higgs Boson: The Nature of Quantum Reality and the Spirit of Copenhagen*. Springer/Nature, New York (2016)

Plotnitsky, A., and Khrennikov, A.: Reality without Realism: On the Ontological and Epistemological Architecture of Quantum Mechanics. Found. Phys., **25** (10), 1269–1300 (2015)

Rovelli, C.: "Incerto Tempore, Incertisque Loci": Can we compute the exact time at which a quantum measurement happens? Found. Phys., **28**, 1031-1043 (1998)

Rovelli, C.: An argument against a realistic interpretation of the wave function. Found. Phys. **46**, 1229-1237 (2016)

Schrödinger, E.: The present situation in quantum mechanics (1935a). In: Wheeler, J.A., Zurek, W.H. (eds) *Quantum Theory and Measurement*, pp. 152–167. Princeton University Press, Princeton (1983)

Schrödinger, E.: Discussion of probability relations between separated systems. Proc. Cambridge Phil. Soc. **31**, 555–563 (1935b)

Schweber, S. S.: *QED and the Men Who Made It: Dyson, Feynman, Schwinger, and Tomonaga*. Princeton University Press; Princeton, NJ (1994)

Silverman, J., Tate, J.: *Rational Points on Elliptic Curves*. Springer, Heidelberg/New York (2015)

Von Neumann, J.: *Mathematical Foundations of Quantum Mechanics* (tr. R. T. Beyer) Princeton University Press, Princeton, NJ, rpt. 1983 (1932)

Wheeler, J. A.: Law without law. In Wheeler J.A., Zurek, W.H. (eds) *Quantum Theory and Measurement*, Princeton University Press, Princeton, NJ., 182-216 (1983)

Wheeler, J.A.: Information, physics, quantum: the search for links. In: Zurek, W.H. (ed) *Complexity, Entropy, and the Physics of Information*. Addison-Wesley, Redwood City, CA (1990)

Wittgenstein, L. *Tractatus Logico-Philosophicus* (tr. C. K. Ogden) London: Routledge, rpt. 1924 (1985)

Zeilinger, A.: A foundational principle for quantum mechanics. Found. Phys. **29** (4), 631-643 (1999)